\documentclass[%
 aip,
 jcp,
 amsmath,amssymb,
 reprint,
]{revtex4-1}
\usepackage{float}
\usepackage[utf8]{inputenc}
\usepackage[T1]{fontenc}
\usepackage{mathptmx}
\usepackage{etoolbox}
\usepackage{graphicx}% Include figure files
\usepackage{dcolumn}% Align table columns on decimal point
\usepackage{bm}% bold math
\usepackage{amsmath}
\usepackage{float}
\usepackage[autostyle]{csquotes}
\usepackage{multirow}
\usepackage{xcolor}
\definecolor{darkgreen}{rgb}{0.0, 0.4, 0.0}
\usepackage[colorlinks=true,linkcolor=blue,citecolor=blue,urlcolor=blue]{hyperref}

\makeatletter
\def\@email#1#2{%
 \endgroup
 \patchcmd{\titleblock@produce}
  {\frontmatter@RRAPformat}
  {\frontmatter@RRAPformat{\produce@RRAP{*#1\href{mailto:#2}{#2}}}\frontmatter@RRAPformat}
  {}{}
}%
\makeatother

\begin{document}

\title[Dynamical and conformational behavior of a polymer in a crowded solution]{Dynamical and conformational behavior of a polymer in a crowded solution}

\author{Setarehalsadat Changizrezaei}
\email{schangiz@uwo.ca}
\affiliation{Department of Physics and Astronomy, The University of Western Ontario, London, Canada}

\author{Colin Denniston}
\email{cdennist@uwo.ca}
\affiliation{Department of Physics and Astronomy, The University of Western Ontario, London, Canada}

\date{\today}

\begin{abstract}
We investigate the structure and dynamics of a polymer in a fluid containing mobile spherical colloidal crowders of radius $R$.  We compare and contrast the behavior with Langevin dynamics (LD) and lattice--Boltzmann molecular dynamics (LBMD), the latter incorporating long-range hydrodynamic interactions. Both the colloid size relative to the monomer radius $r$ and the volume fraction $\phi$ are varied to determine how crowding modifies polymer behavior.
Increasing volume fraction induces polymer compaction, with the mechanism strongly dependent on the size ratio $R/r$. 
Small colloids primarily  modify the short-wavelength polymer conformation,
causing self-avoiding-walk-like behavior to persist to shorter length scales, whereas large colloids reduce the effective long-wavelength Flory exponent, indicating degraded solvent quality consistent with a confinement-blob picture.
Polymer diffusion exhibits distinct behavior in LD and LBMD. In LD, diffusion decreases rapidly and depends strongly on $R/r$; a phenomenological scaling involving $\ln(1+R/r)$ captures this size dependence, and additional scaling with $R_g$ reduces scatter, indicating polymer-scale correlations induced by crowding. In contrast, LBMD diffusion follows an effective-medium–like exponential dependence on concentration, governed by hydrodynamic coupling. Rouse-mode analysis identifies three regimes: scaling breakdown at low volume fraction, Zimm-like behavior at intermediate density in both LD and LB, and at high density hydrodynamic screening in LB with confinement-dominated dynamics in LD.
\end{abstract}

\maketitle

%%%%%%%%%%%%%%%%%%%%%%%%%%%%%%%%%%%%%%%%%%%%%%%%%%%%%%%%%

\section{Introduction}
The effects of nanoparticle (NP) crowders on polymer conformation and dynamics
have been examined in a range of studies~\cite{10.1063/1.1874852, Li_2015, 10.1063/1.3105336, C0CP02952A, doi:10.1021/acs.macromol.0c00158}.  Nanoparticle-crowded environments are particularly important in biological
systems~\cite{GOODSELL1991203, ELLIS2001597}; for instance, living cells often contain a variety of substances like chromatin, actins, nuclei, cytoskeletons, etc~\cite{Saxton_2007, Weiss_2004,Jeon_2011, Di_Rienzo_2014}. Dynamic properties of polymers such as RNA and proteins are influenced by crowders and it has been observed that the way polymers behave in living cells differs from their behavior in a simple aqueous solution~\cite{PhysRevE.94.022614}. One striking example is DNA condensation, also known as the coil-globule transition. In this phenomenon, the size of DNA reduces from a swollen coil in dilute solutions to a compact globule, mediated by interactions with both electrostatically charged and uncharged proteins.\cite{C3SM51214B, doi:10.1021/nl203114f, doi:10.1021/jp2124907}. 

The freely jointed chain (FJC)~\cite{Smith1992-xa} and the worm-like chain (WLC)~\cite{https://doi.org/10.1002/recl.19490681203,Bustamante1994-lc} models are widely used to study synthetic ~\cite{doi:10.1021/la980853t,doi:10.1021/ma981245n} and natural polymers~\cite{doi:10.1126/science.276.5315.1109} such as
single-stranded DNA~\cite{doi:10.1126/science.271.5250.795}. The WLC accounts for the bending stiffness of the chain which leads to orientational correlations over a characteristic length scale known as the persistence length ($L_p$). The overall size of the full chain configuration is then characterized by its radius of gyration $(R_{g})$.  The dynamics of the polymer depends significantly on the nature of its environment~\cite{Doi}, including diffusion and scaling properties.  Early simulations of single polymers in solution \cite{Ceperley1981} relied on Langevin dynamics which do not include hydrodynamics leading to diffusion depending primarily on the polymer length.  In contrast, models and simulations that include hydrodynamics~\cite{A.Malevanets_2000,10.1063/1.465445,10.1063/1.1855876,Ollila2011-wg} lead to diffusion depending on $R_g$.  

Polymers in confined environments are often well described by de Gennes so-called "blob" theory \cite{deGennes77, deGennes77b}.  This works well in environments where there is a reasonably well-defined length scale for the confinement which set the size of a blob.  Below the blob length scale the polymer acts like an unconfined polymer while for length scales much longer than the blob scale, the polymer can be viewed as a polymer of blobs whose confirmation depends signifantly on the details of the confinement.  Similar models have also been constructed to describe long polymers in a crowded solution of short polymers \cite{JoannyA, JoannyB, Adler80}.  The dynamics of polymers in crowded environments is not as well understood and is an active area of investigation.

A number of studies examined polymer dynamics in the presense of immobile nanoparticle crowders using off-lattice Monte Carlo \cite{C7CP05514E} or Langevin simulations \cite{doi:10.1021/acs.macromol.2c01786,D2SM00974A} (implicit solvent, no hydrodynamics).  These studies typically see changes in the diffusive dynamics depending on the length of the polymer and in particular if it is long enough to interact with multiple versus a single crowder particle.  Shorter polymers exhibit discrete hops from NP to NP leading to subdiffusive behaviour.  This dampens out and disappears for longer polymers.  An active polymer \cite{D2SM00974A} can also transition between non-local diffusive states and localized states depending on the strength of the active force. The level of disorder of the crowder configuration can also affect the dynamics \cite{doi:10.1021/acs.macromol.2c01786}.
  
The combined impact of mobile crowders and confinement by walls has also been studied with Langevin dynamics \cite{D0SM01847C, 10.1063/5.0054797,10.1063/1.4919650}.  The crowders can lead to effective interactions \cite{D0SM01847C} and enhanced adsorption to a wall \cite{D0SM01847C,10.1063/5.0054797}.  The combinination of crowder and wall interactions can lead to the polymer size swelling under weak confinement.  In contrast, strong confinement and high volume fractions of crowders can cause the polymer to decrease in size \cite{10.1063/1.4919650}.
  
Echeverria et al.~\cite{10.1063/1.3319672} studied both the conformational and diffusional dynamics of globular polymers in a poor solvent containing fixed spherical obstacles. These properties are examined as a function of the obstacle volume fraction, obstacle size, and the length of the polymer chain. Both hydrodynamic interactions among the polymer beads and intermolecular forces involving the solvent molecules are considered. They observed a nonmonotonic change in the radius of gyration over time, and notably, the time scale for collapse is significantly longer when obstacles are present compared to simple solutions without obstacles. Hydrodynamic interactions play a crucial role at low obstacle volume fractions, but their significance diminishes at high volume fractions due to screening effects. In systems with high obstacle volume fractions, large polymer chains tend to adopt blob-like conformations. This conformational behavior arises from the trapping of segments of the chain within the void spaces among the obstacles. The movement of the globular polymer chain through obstacles exhibits subdiffusive behavior on intermediate time scales. During this period, the dynamics involve the exploration of the local structure within the heterogeneous environment.

In our work, we investigate the conformational and dynamical behavior of a polymer in fluid crowded with mobile nanoparticles using hybrid Lattice Boltzman-Molecular dynamics simulations, which explicitly includes long-range hydrodynamic interactions. The main contribution of this work is a systematic investigation of the role of hydrodynamic interactions in the behavior of a polymer immersed in a crowded solution of mobile colloids. By directly comparing Lattice-Boltzmann (LBMD) simulations with traditional Langevin dynamics, we provide new quantitative insights into how these often-overlooked, fluid-mediated forces influence the polymer’s structure and dynamics. 

 \begin{figure}[t]
	\begin{center}
		\includegraphics[trim={75 0 0 0},scale=0.22]{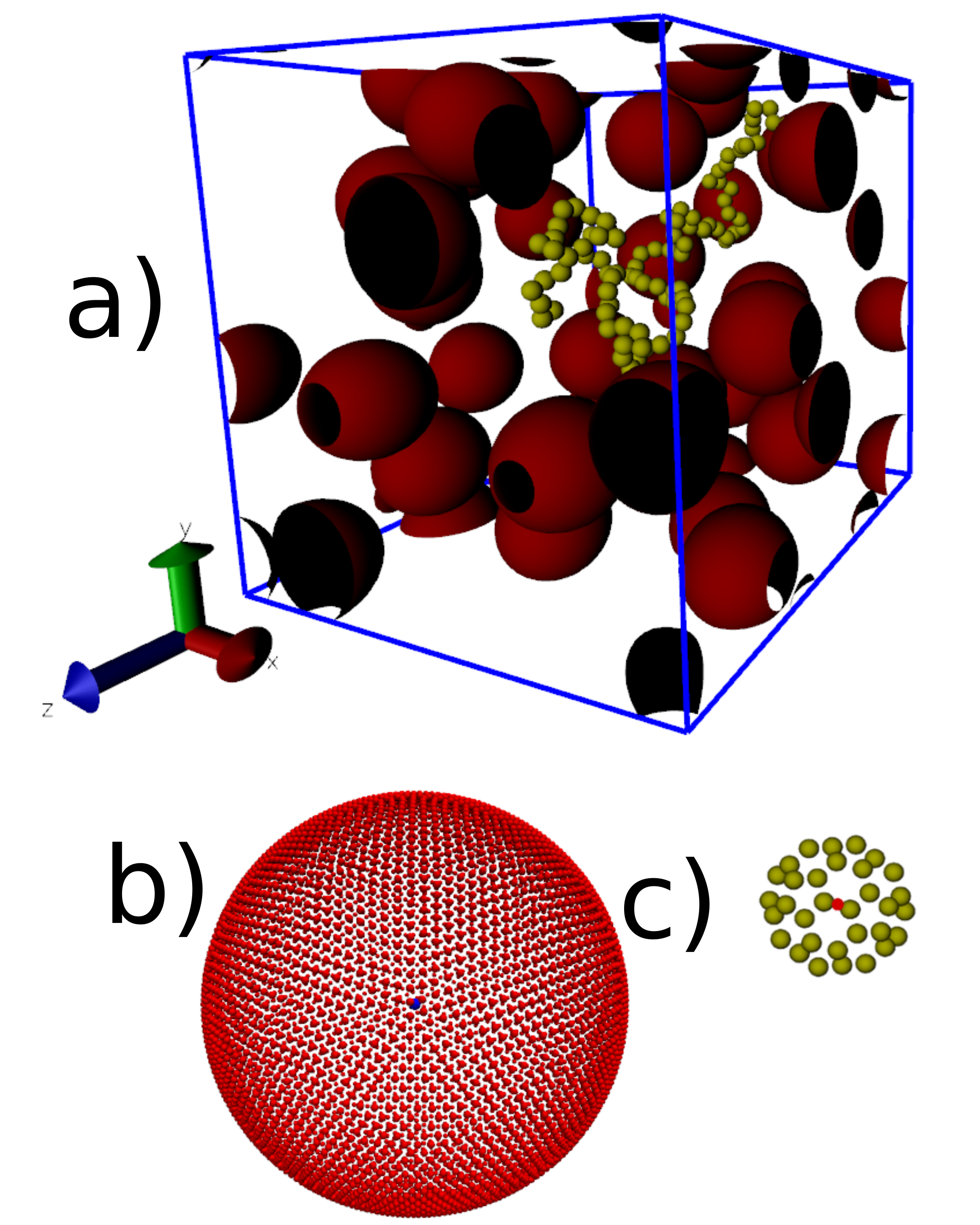}
		\caption{A schematic representation of the simulation system. (a) A polymer chain (yellow) is shown within the simulation box, surrounded by a crowded solution of mobile spherical colloids (red). (b) Each colloid consists of a central atom and a spherical surface shell. (c) Each monomer is similarly composed of a central atom and a smaller surface shell. Direct particle interactions, such as the FENE bond potential and Lennard-Jones repulsion, are applied between the central atoms, while the shell nodes interact with the Lattice-Boltzmann fluid to capture long-range hydrodynamic effects.
		}
		\label{box}
	\end{center}
\end{figure}

\section{Modeling}
The LBMD approach combines molecular dynamics (MD) for particle modeling and the lattice-Boltzmann (LB) method for fluid representation. In Figure 1(a), we present a schematic representation of our system. Periodic boundary conditions were used in the x-, and y- and z-directions.
All the simulations were performed using LAMMPS~\cite{Thompson2022-lp}. The linear polymer chain is comprised of 96 or 192 monomers, with interactions between non-consecutive monomers along the chain modeled through Lennard-Jones (LJ) interactions:
\begin{equation}
	U_{LJ}(r;\sigma,\epsilon) = 4\epsilon \left[ \left(\frac{\sigma}{r} \right)^{12} - \left(\frac{\sigma}{r} \right)^6 + A
	\right] \Theta(r_c-r)
\end{equation}
where $r=|\vec{r}_i-\vec{r}_j|$ is the distance between non-adjacent monomers, $\sigma$ and $\epsilon$ are the LJ distance and energy scales, $r_c$ is the cutoff, $\Theta$ is the Heaviside step function, and $A$ is chosen so that $U_{LJ}(r_c)=0$.  Here we use a purely repulsive LJ (the Weeks-Chandler-Andersen (WCA) potential~\cite{Weeks1971-ra}) where $r_c=2^\frac{1}{6}\sigma$ (and $A=1/4$), and used $\sigma\!=\!4.0\,\mathrm{nm}$, and $\epsilon = 1.0\,\mathrm{ag} (\mathrm{nm})^2 (\mathrm{ns})^{-2}$.

The sequential connection of monomers along the chain was implemented using Kremer and Grest's finitely extensible non-linear elastic (FENE)~\cite{46ref_p} bond potential 
\begin{equation}
	\label{FENE}
	U(r)= -\frac{1}{2} KR_0^2\log\left(1-\frac{r^2}{R_0^2}\right) + U_{LJ}(r;\sigma,\epsilon_m).
\end{equation}
where $r=|\vec{r}_i-\vec{r}_{i\pm 1}|$ is the distance between adjacent monomers, 
$K=60\, \mathrm{ag}\, \mathrm{ns}^{-2}$, and $R_0 = 6\, \mathrm{nm}$ is the maximum bond extension.  The $LJ$ part of the potential is similar to that for non-adjacent monomers except that $\epsilon_m = 4.14\,\mathrm{ag} (\mathrm{nm})^2 (\mathrm{ns})^{-2}$.

An additional WCA potential was applied between the monomers of the polymer and the colloids to prevent overlap. For that, we chose $\epsilon_\mathrm{ms} = 1.0\,\mathrm{ag} (\mathrm{nm})^2 (\mathrm{ns})^{-2}$ and $\sigma_\mathrm{ms} = \sigma/2 + R $ for the interaction between the monomers and spherical colloids of radius $R$. These parameters are set large enough to ensure that no overlaps occur. A repulsive LJ potential  was applied between the central atoms of the colloidal spheres. We chose $\epsilon_\mathrm{spheres} = 1.0 \, \mathrm{ag} (\mathrm{nm})^2 (\mathrm{ns})^{-2}$ and $\sigma_\mathrm{sphere}= 2 R$ for interaction between the spheres.

To include hydrodynamic interactions between the particles and the solvent, we
used a hybrid lattice-Boltzmann molecular-dynamics (LBMD) approach, as implemented through the LB fluid package
in LAMMPS~\cite{27ref_p,Ollila2011-wg,DENNISTON2022108318,Thompson2022-lp}. In this
method, the polymer monomers and colloidal crowders are evolved using molecular
dynamics, while the solvent is represented separately by a lattice-Boltzmann
fluid. We employed the $D_3Q_{15}$ lattice-Boltzmann model, which uses 15
discrete velocities on a three-dimensional cubic lattice~\cite{29ref_p,
32ref_p,33ref_p}. The fluid dynamics are obtained by solving a discretized
form of the Boltzmann equation.

The molecular-dynamics particles exchange momentum with the surrounding LB
fluid, and the fluid transports this momentum throughout the simulation box.
Consequently, the motion of one particle can affect other particles through
the solvent, giving rise to long-range hydrodynamic interactions. 
In this approach, thermal fluctuations are introduced through the LB fluid,
which acts as a momentum-conserving thermal bath for the molecular-dynamics
particles~\cite{Ollila2011-wg}. Thus, both hydrodynamic interactions and thermal fluctuations are
included in the LBMD simulations. This differs from the Langevin-dynamics
simulations used for comparison, in which the solvent is implicit and
long-range hydrodynamic interactions are absent.

Mass and momentum conservation are captured by the following expressions:
\begin{equation}
	\begin{split}
		&\partial_t \rho + \partial_\alpha\left(\rho u_\alpha\right)=0  \\
		& \partial_t \rho + \partial_\beta(\rho u_\alpha u_\beta) = -\partial_\alpha P_{\alpha \beta} + F_\alpha \\
		&\!\! +\! \partial_\beta [\eta (\partial_\alpha u_\beta \!+\! \partial_\beta u_\alpha \!-\! \frac{2}{3} \partial_\gamma u_\gamma \delta_{\alpha \beta} ) \!+\! \zeta \partial_\gamma u_\gamma \delta_{\alpha \beta}  ] \!+\! s_{\alpha\beta}  ,
	\end{split}
	\label{density}
\end{equation} 
where $\eta$ and $\zeta$ are the shear and bulk viscosities, $P_{\alpha \beta}$ is the fluid pressure, and $F_\alpha$ is the force density exerted by the molecular-dynamics particles on the LB fluid. The shear viscosity in the model is $\eta = \rho \tau v_c^2/3$, where $v_c = \frac{\Delta x}{\Delta t}$ is a lattice velocity, and $\zeta = \eta (5/3 - 3v_s^2/v_c^2)$~\cite{PhysRevLett.75.830}. The speed of sound $v_s^2=v_c^2/3$.
In our system, the lattice spacing is $1\,\mathrm{nm}$ and time step is $0.0003\,\mathrm{ns}$. In order to speed up the diffusive dynamics~\cite{C3SM27410A,Ollila2011-wg}, we set the viscosity of the fluid at a value that is 1/60th of that of water at a temperature of $T=300$\,K.  
%The lattice Boltzmann simulations were performed in a periodic box of  
This reduced viscosity increases the absolute diffusion coefficients and
shortens the relaxation times, making the long-time motion of the polymer
computationally accessible. In the hydrodynamic regime, the polymer
center-of-mass diffusion coefficient scales approximately as
$D_{\mathrm{cm}}\sim k_{\mathrm{B}}T/(\eta R_g)$, so reducing $\eta$
accelerates the dynamics.

In the LBMD simulations, each polymer monomer and colloid is represented as a
composite molecular-dynamics particle consisting of a central atom surrounded
by a spherical shell of surface nodes, as illustrated in
Figure~\ref{box}. The central atoms determine the direct molecular
interactions. In particular, the FENE bond potential acts between bonded
polymer atoms, while the Lennard-Jones interactions act between the relevant
central atoms.  The surface nodes provide the coupling between the molecular-dynamics
particles and the LB fluid. Through this coupling, momentum is transferred
between each particle and the surrounding solvent. The momentum is then
transported through the LB fluid, producing the long-range hydrodynamic
interactions between the polymer and colloids. This composite-particle
construction therefore separates the direct bonded and excluded-volume
interactions, which are applied through the central atoms, from the
hydrodynamic coupling, which is applied through the surface nodes. The detailed information can be found in the references~\cite{44ref_p,27ref_p,39ref_p}.

The spherical shell also provides a well-defined hydrodynamic radius for each
particle~\cite{39ref_p}. Because the hydrodynamic size includes
both the surface mesh and the intrinsic contribution associated with the
particle--fluid coupling, the radius of the surface mesh is chosen to be
slightly smaller than the intended colloidal radius, with the correction
determined by the LB lattice spacing $\Delta x$~\cite{39ref_p}. This model has
previously been used successfully to study a range of colloidal
systems~\cite{C4SM01812E,Ollila_Ala-Nissila_Denniston_2012,
PhysRevE.87.050302} and polymer systems in which hydrodynamic interactions are
important~\cite{D4SM00761A,C8SM01445K,Ollila2011-wg,C3SM27410A,
Ollila2014-xg,D0SM01045F,doi:10.1021/acs.biomac.3c00473}.

In order to elucidate the role of hydrodynamics, for comparison we also performed Langevin simulations (LD), in which the Langevin equation was integrated forward in time using the velocity-Verlet algorithm in the LAMMPS, and it models an interaction with a background implicit solvent. At a mesoscopic timescale, The Langevin equation can be used to describe the motion of a Brownian particle:
\begin{equation}
    m_i \dot{\vec{v}}_i
    =-\nabla_i U-\gamma_i \vec{v}_i+\vec{F}_{r,i},
    \label{LD}
\end{equation}
where $m_i$, $\vec{r}_i$, and $\vec{v}_i$ are the mass, position, and velocity
of particle $i$, respectively, and $U$ is the total interaction potential.
The friction coefficient is related to the damping time by
$\gamma_i=m_i/\tau_{\mathrm{damp},i}$, and $\vec{F}_{r,i}$ is a zero-mean
random thermal force satisfying the fluctuation--dissipation relation. The
magnitude of this force scales as
$\sqrt{k_{\mathrm B}T m_i/(\Delta t\,\tau_{\mathrm{damp},i})}$.

In the Langevin simulations (LD), we used a time step of
$\Delta t=0.007\,\mathrm{ns}$ and a temperature in LJ units of $T^*=2.76$,
corresponding to $T=300\,\mathrm{K}$ in real units. The damping times were
$20\tau_{\mathrm{LJ}}$ for the polymer monomers and
$10\tau_{\mathrm{LJ}}$ for the crowders, where the Lennard-Jones time scale is defined as
$\tau_{\mathrm{LJ}}=\sigma\sqrt{2m_{\mathrm m}/\epsilon}$, where $\sigma$ and
$\epsilon$ are the Lennard-Jones parameters used for the polymer in both LD and LBMD simulations. These values were chosen to control
the relative rates of particle diffusion and relaxation in the
implicit-solvent simulations rather than to reproduce the exact size
dependence predicted by the Stokes relation, $\gamma=6\pi\eta R$. In
particular, the crowders remain sufficiently mobile to rearrange around the
polymer rather than behaving as nearly fixed obstacles.
The polymer monomer mass was $m_{\mathrm m}=0.223\,\mathrm{ag}$, while the
crowder masses were determined from their radii and the mass
density (same as fluid). 
%}

All simulations were performed in a periodic cubic box of dimensions
$215\,\mathrm{nm} \times 215\,\mathrm{nm} \times 215\,\mathrm{nm}$.
For each colloid radius and volume fraction, the colloids were initially
placed on an FCC lattice with lattice constant just small enough to obtain the desired $\phi$ or slightly higher.  Particles were removed as needed to obtain the desired value of $\phi$. The polymer was then inserted into the available
space, and the initial configuration was checked to ensure that it did not
overlap with any colloids.

Each system was equilibrated for at least $10^{6}$ simulation steps before
any measurements were taken, and all data from this equilibration period were
discarded. For the lattice-Boltzmann simulations, this corresponds to
$300\,\mathrm{ns}$, whereas for the Langevin-dynamics simulations it
corresponds to $7\,\mu\mathrm{s}$. The equilibration time was assessed by
monitoring both the polymer and colloidal structures. In particular, the
polymer radius of gyration, $R_g$, was plotted as a function of time, and its
running average was examined. After $10^{6}$ steps, $R_g$ showed no continuing
systematic drift and instead fluctuated around a stable post-equilibration
mean (a sample plot of $R_g$ versus time is shown in the Supplementary material).

Equilibration of the colloids was also examined by calculating the
colloid--colloid radial distribution function, $g(r)$, over consecutive,
nonoverlapping time windows. The height of the first peak,
$g(r_{\mathrm{peak}})$, showed no systematic drift after $10^{6}$ steps and
fluctuated around a stable mean (a sample plot of $g(r_{\mathrm{peak}})$ versus time is shown in the Supplementary material). The consistent behavior of $R_g$ and
$g(r_{\mathrm{peak}})$ indicates that both the polymer and the local colloidal
structure had reached steady states before production sampling began.

After equilibration, configurations and observables were recorded every 500
simulation steps. The reported radius of gyration and other structural and
dynamical quantities were obtained by averaging over the production portion
of each trajectory. The minimum total simulation length was
$10{,}000{,}000$ steps, leaving at least $9\times10^{6}$ steps for production
sampling after the equilibration period.

\section{Results}

\begin{figure}[tb]
		\includegraphics[trim= 0 0 0 0cm,scale=0.31]{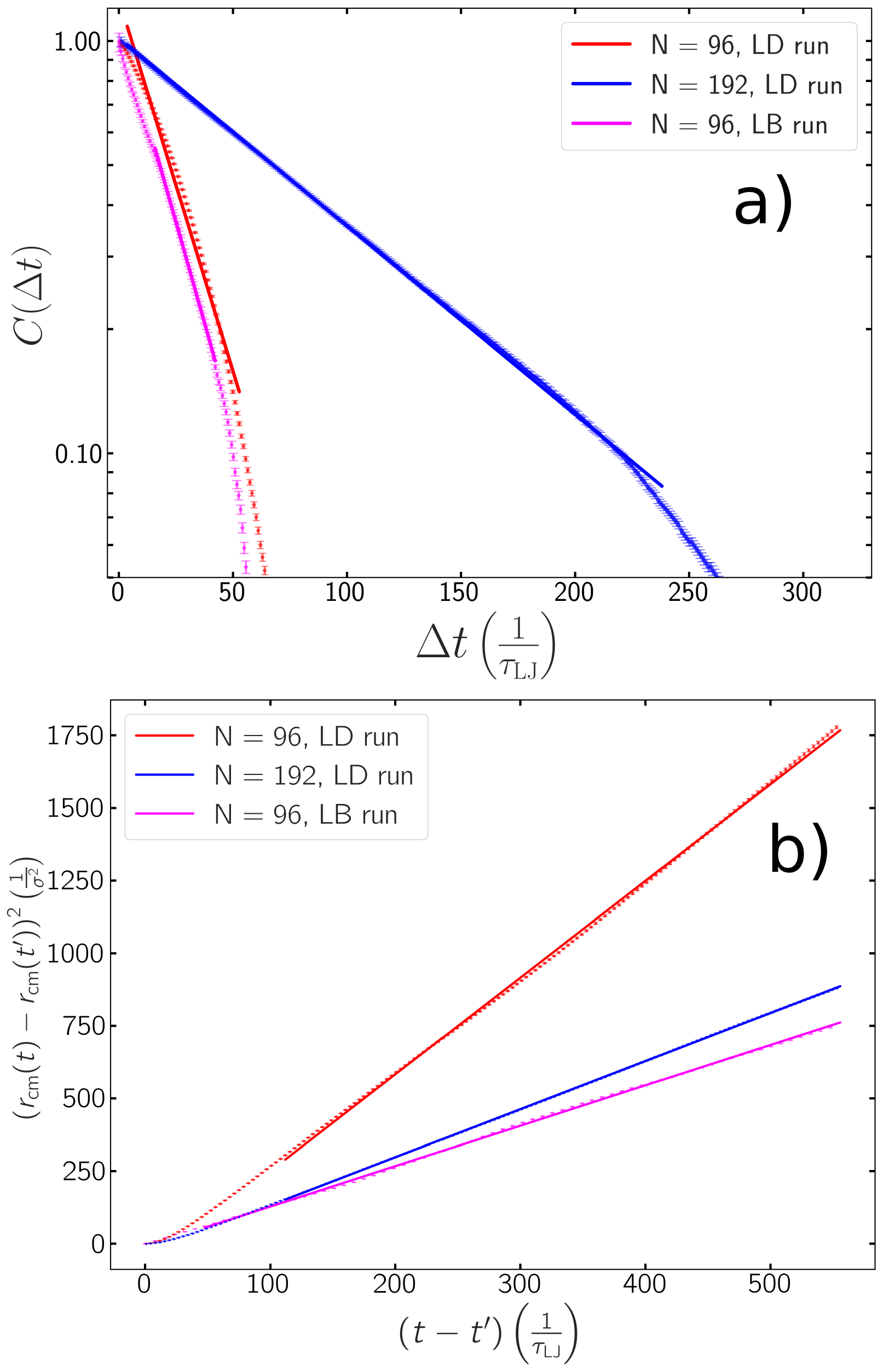}
		\caption{a) The time autocorrelation of the end-to-end distance of the polymer as a function of lag time. The line corresponds to the fit of the time autocorrelation, which is used to determine the relaxation time of the polymer.  b) Time evolution of Mean Squared Displacement (MSD), $\langle \left( \bm{r}(t) -\bm{r}(t^\prime)\right)^2\rangle$ in the absence of colloidal particles within the system.  The lines correspond to the fits used to determine the diffusion constant $D$.
		}
		\label{C_t_MSD}

\end{figure}

\subsection{Polymer Characteristics}

We begin by characterizing a single polymer in the absence of colloidal
crowders. The parameters of the Langevin and lattice-Boltzmann simulations
were chosen so that the static equilibrium properties should agree within
statistical uncertainty. However, their dynamical properties differ because
hydrodynamic interactions are included in the LB simulations but are absent
from the Langevin simulations, which use an implicit solvent. Langevin
simulations were performed for polymers containing $N=96$ and $N=192$
monomers, whereas the more computationally expensive LB simulations were
performed only for $N=96$.

A key length scale for the polymer in any system is the radius of gyration $(R_{g})$ of the chain which is defined as,
\begin{equation}
	R_g = \sqrt{\frac{1}{N}\sum_i ({\mathbf r}_i-{\mathbf r}_{com})^2}
\end{equation}
where ${\mathbf r}_i$ is the position of monomer $i$, and ${\mathbf r}_{com}$ is the centre of mass of the chain.  The equilibrium value in solution with no crowders $R_{g0}$ for a polymer consisting of 96 monomers was measured to be $6.6\sigma$ for the Langevin case and $6.7\sigma$ for
the LB case, and it increased to $10.0\sigma$ when the number of monomers in the polymer chain was doubled to 192.  The measured equilibrium radii are consistent with standard models~\cite{Doi}
\begin{equation}
\label{RgScalingPhi0}
    R_{g0} \approx (0.5b)N^\nu .
\end{equation}
where $b=0.9\sigma$ is related to the mean bond length or effective monomer size, $N$ is the number of monomers, and $\nu=0.59$ is the well-known Flory exponent of a self-avoiding walk.

The polymer's relaxation time is determined by evaluating the time autocorrelation of its end-to-end distance. This autocorrelation function typically follows an exponential decay, described by $\exp(-t/\tau)$, where $\tau$ denotes the polymer's relaxation time~\cite{Doi}.  
Figure~\ref{C_t_MSD}a illustrates the computed auto-correlation function. For the Langevin simulations, the relaxation time is found to be approximately $\tau = 24.07 \pm 1.26\tau_{LJ}$ for $N=96$ and $\tau = 95.08 \pm 0.52\tau_{LJ}$ for $N=192$.  For the LB case, we found $\tau = 22.18 \pm 0.94\tau_{LJ}$.  For an ideal chain (undergoing a random walk) in the absence of hydrodynamics, the longest relaxation time scales as $\tau_N \sim N^2$ \cite{Doi}.  For self-avoiding chains (like we have here), a similar scaling analysis leads to a relaxation time, again ignoring hydrodynamic effects, of $\tau_N \sim N^{1+2\nu}$, with $\nu=0.59$ \cite{Ceperley1981}.  This scaling would imply that for the Langevin cases, the ratio of the $\tau$ for $N=96$ and $N=192$ should be 0.22, very close to the value of 0.21 that we measure.  LB includes hydrodynamics so we do not expect the same relaxation time. The similarity of the values for $\tau$ here are somewhat coincidental.  In any case, for the LB we only have one chain length and we will examine the dynamic scaling of the Rouse modes for the chain in the last section of the paper. 

We measured the mean-squared displacement (MSD) of the center of mass of the polymer $\langle \left( \bm{r_{cm}}(t) -\bm{r_{cm}}(t^\prime)\right)^2\rangle$ as a function of time to find the diffusion coefficient of the polymer. The MSD is calculated over lag time exceeding the relaxation time, ensuring the system is in the linear regime. Fig.~\ref{C_t_MSD}b illustrates the temporal changes in $\langle \left( \bm{r_{cm}}(t) -\bm{r_{cm}}(t^\prime)\right)^2\rangle$ for our polymers in the absence of colloidal particles within the system. Evidently, $\langle \left( \bm{r_{cm}}(t) -\bm{r_{cm}}(t^\prime)\right)^2\rangle$ exhibits a linear increase with time $(t-t^\prime)$, indicative of normal diffusion.  Subsequently, the diffusion coefficient (D) of the polymer can be derived through the following relation:
\begin{equation}
	\begin{split}
	&D = \lim_{\left|t-t^\prime \right| \to \infty}\frac{\langle \left( \bm{r_{cm}}(t) -\bm{r_{cm}}(t^\prime)\right)^2\rangle}{6(t-t^\prime)}
	\end{split}
\end{equation}
The slope of $\langle \left( \bm{r_{cm}}(t) -\bm{r_{cm}}(t^\prime)\right)^2\rangle$ versus $(t-t^\prime)$ is equal to $6D$ and therefore $D_0$(the diffusion coefficient of chain in the bulk) can be obtained from the linear fit of MSD as a function of $(t-t^\prime)$.   For the Langevin simulations this gives 0.55${\sigma}^2/\tau_{LJ}$ for $N=96$ and 0.27${\sigma}^2/\tau_{LJ}$ for $N=192$ from fits to the data shown in Fig.~\ref{C_t_MSD}b. A fluctuation-dissipation relation dictates that the diffusion constant is inversely related to the drag. For the Langevin simulations, where each monomer experiences the drag and fluctuations independently, this gives \cite{Doi}
\begin{equation}
	D_{LG}= \frac{k_B T}{N \gamma}.
	\label{DLG}
\end{equation}
The expected scaling, $D\sim 1/N$ is observed as the diffusion constant for the $N=192$ case is very close to half that of the $N=96$ case.  The presence of hydrodynamics significantly changes this scaling as it induces correlated motions that result in a drag force proportional to the size of the polymer $R_g$, resulting in ~\cite{A.Malevanets_2000, 10.1063/1.465445,10.1063/1.1855876}
\begin{equation}
	D_{\text{cm}} = \frac{k_BT}{6\pi \eta}\frac{A}{R_g},
		% D_{\text{cm}} = \frac{k_BT}{6\pi \eta}(\frac{A}{R_g}-\frac{B}{L}),
		\label{DLB}
\end{equation}
where $A$ is a dimensionless constant~\cite{Ollila2011-wg} of order one.  There can also be finite-size effects in a periodic box due to the polymer "seeing itself" through the periodic boundaries. 
For our 96-mer in the lattice Boltzmann fluid this is measured to be $0.23{\sigma}^2/\tau_{LJ}$, a smaller but similar value compared to the corresponding outcomes in LD simulations for polymer lengths N=192 and N=96. Again, in this case the similarity is somewhat conicidental and is due to our choice of drag coefficient in the Langevin simulations.

\begin{figure}[tb]
	\begin{center}
		\includegraphics[trim= 0 0 0 0cm,scale=0.50]{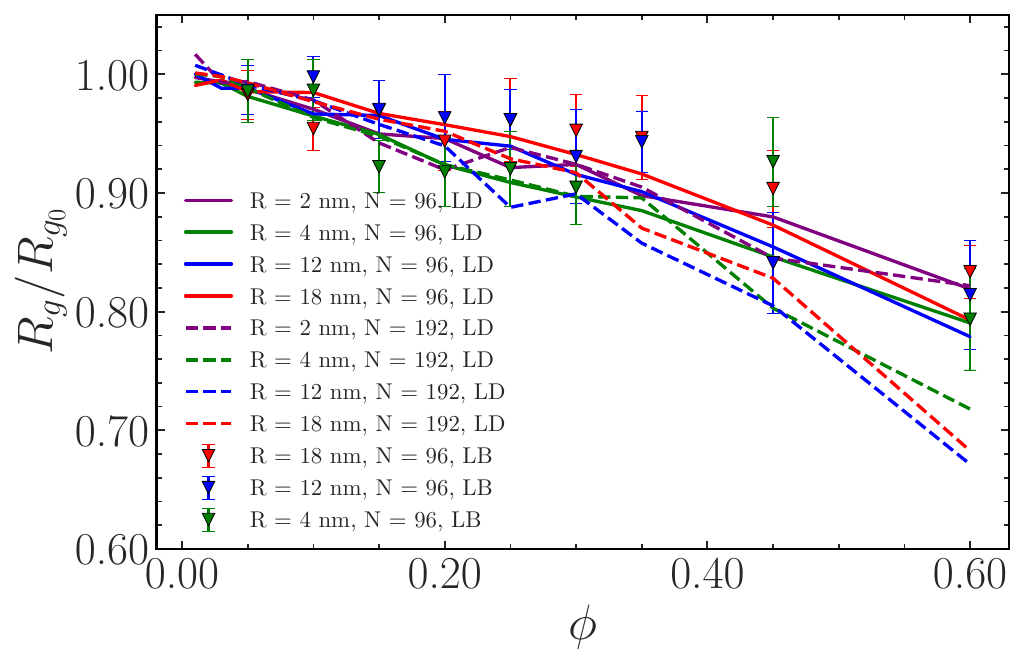}
		\caption{The polymer radius of gyration normalized by its value in free
solution, $R_g/R_{g0}$, as a function of the colloid volume fraction $\phi$.
Solid lines show the $N=96$ lattice-Boltzmann results, dotted lines show the
$N=96$ Langevin-dynamics results, and dashed lines show the $N=192$
Langevin-dynamics results.}
		\label{RgvsPhi}
	\end{center}
\end{figure}

\begin{figure*}[tb]

		\includegraphics[trim= 60 0 0 0cm,scale=0.115]{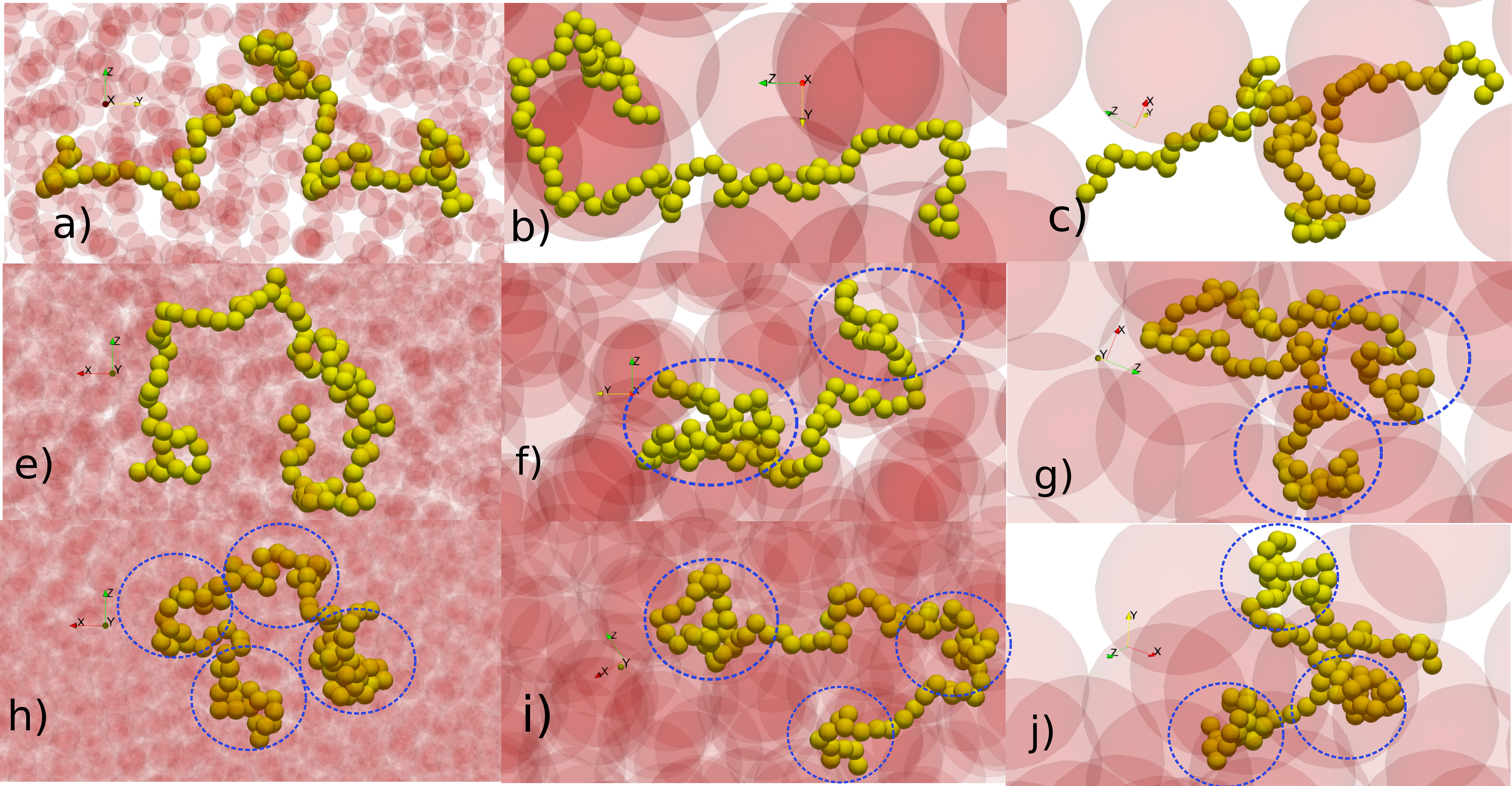}
		\caption{The polymer and colloid conformations in lattice Boltzmann simulations at low volume fraction of 0.10 for a) R=4nm, b) R=12nm, c) R=18nm, at intermediate volume fraction of 0.35 for d) R=4nm, e) R=12nm, f) R=18nm, and at high volume fraction of 0.60 for g) R=4nm, h) R=12nm, and i) R=18nm. 
		Monomer clumps, or clusters, that will potentially give rise to de Gennes "blob"-like polymer behavior are indicated by the blue dashed regions. (Multimedia available online for (f) and (i))}
		\label{LB_pol}

\end{figure*}

\subsection{Polymer with Crowders: Static Equilibrium Properties} \label{ssec:Static Equilibrium Properties}
We now examine the influence of mobile colloidal crowders on the dynamics and
conformation of the polymer. We consider crowders with radii
$R=2\,\mathrm{nm}$, $4\,\mathrm{nm}$, $12\,\mathrm{nm}$, and
$18\,\mathrm{nm}$, with colloid volume fractions ranging from 0.05 to 0.60.

We first examine the radius of gyration of the polymer as a function of the volume fraction $\phi$ of the colloids. In Figure~\ref{RgvsPhi}, radius of gyration of the chain scaled by $R_{g_0}$, which depended only on $N$, is plotted as a function of $\phi$. Static equilibrium properties of the chain such as $R_g$ are expected to be independent of the dynamics (Langevin or lattice Boltzmann) and this is seen in the plot:  while the lattice Boltzmann data are noisier due to the necessarily shorter time duration runs (due to increased computational cost), the results agree reasonably between the LB and langevin simulations.  As the volume fraction increases, the radius of gyration decreases for polymers of both molecular weights and all colloid sizes, showing that the chains become more confined and adopt more compact structures.

At low to intermediate volume fractions, the radius of gyration for all systems shows similar patterns. However, at an intermediate volume fraction, the data start to diverge.  The higher molecular weight polymer showing a larger decrease in radius of gyration for all but the smallest colloidal size. 
In addition, as the colloid size increases, the spread between the lines grows. 
The observed reduction in radius of gyration for N=192, particularly for larger colloidal sizes, might suggest a tendency to form more compact effective "blobs" compared to the shorter polymers. 

Figure~\ref{LB_pol} (Multimedia available online) presents snapshots of chain conformations from lattice Boltzmann simulations (Langevin looks similar, see supplementary for different colloid radii (\( R = 4 \, \text{nm} \), \( R = 12 \, \text{nm} \), and \( R = 18 \, \text{nm} \)) at low (\( \phi = 0.10 \)), intermediate (\( \phi = 0.35 \)), and high (\( \phi = 0.60 \)) volume fractions.  
At low volume fraction (first row of figure~\ref{LB_pol}), the configuration of the polymer chain is not that different from a free chain, although there are short sections where we can see some clustering from the chain's random walk being impeded by the presence of the colloids.  
At an intermediate volume fraction, the denser clusters appear more varied in size with several larger clusters for \( R = 12 \, \text{nm} \) and \( R = 18 \, \text{nm} \) comparing to the more uniform \( R = 4 \, \text{nm} \).  This is related to the fact that the individual "cavities" or holes in the colloidal suspension are much larger for the larger colloids, even at the same overall volume fraction.  This increased spacing between colloids allows more room for the chain to form compact, larger clusters (indicated by blue dashed regions in the figure). At a high volume fraction, the polymer becomes further compacted.  This compaction is fairly uniform across length scales for the smaller colloids (\( R = 4 \, \text{nm} \)) while in the larger colloids we see larger clusters in the cavities of the colloidal suspension connected by thinner (less compacted) polymer strands.  However, determining differences in polymer configuration based solely on visual observations of monomer arrangements is challenging. To accurately assess these differences, we next quantify the polymer structure by examining the structure factor of colloids and monomers.

 \subsubsection{Static polymer scaling}
 The radius of gyration, \( R_g \), discussed above is commonly used to characterize the size of a polymer. In solvents without crowders it scales with the degree of polymerization, \( N \), following the relation \( \langle R_g \rangle \sim b N^\nu \), where \( \nu \) is the Flory exponent reflecting the influence of excluded volume interactions. A polymer in a good solvent typically follows a self-avoiding random walk in three dimensions with \( \nu \approx 0.5877 \)~\cite{Li1995}.  A polymer in a poor solvent tends to form more compact configurations characterized by  \( \nu \approx 0.33 \) while a neutral solvent is in-between with \( \nu \approx 0.5 \).  As a result, when we see a reduction in $R_g$ with increasing concentration of colloidal crowders there are a number of possible situations that could be going on.  The polymer could be adopting more compact configurations similar to that found in a poor solvent that decreases the scaling exponent $\nu$ at long wavelengths.  Another possibility is that colloids could impact the chain configurations at shorter wavelengths making it appear less stiff, similar to what you would get if the effective persistence length were getting shorter.  There could also be a mix of effects occuring between short and long wavelengths. 
 
 The Flory exponent \( \nu \) can be determined from the averaged static structure factor \( S(k) \), which is defined in terms of the monomer density function, 
 $$\rho(r) = \sum_{m}^{} \delta(r - r_m),$$ and its Fourier transform, $$	\hat{\rho}(k) = \mathcal{F}(\rho(r)) = \sum_{m}^{} \exp(i k \cdot r_m),$$ where the sums are over monomers and $r_m$ is the position of monomer $m$. The static structure factor is then given by  
 \begin{equation}
 	S(k) = \frac{1}{N} \int \hat{\rho}(k) \hat{\rho}^*(k) \, d\Omega,
 \end{equation}  
 which, upon substitution, simplifies to  
 \begin{equation}
 	S(k) = \frac{1}{N} \sum_{m,n} \frac{\sin(k |r_m - r_n|)}{k |r_m - r_n|}.
 \end{equation}  
 For an appropriate range of wave vectors \( k \), the static structure factor is expected to follow a characteristic scaling relation:  
 \begin{equation}
 	S(k) \sim k^{-1/\nu}.
 \end{equation}  
As a result, one expects 
\begin{equation}
	\frac{d \log S(k)}{d \log k}  = -1/\nu.
\end{equation}

 \begin{figure}[tb]
 		\includegraphics[trim= 7 0 0 0,scale=0.255]{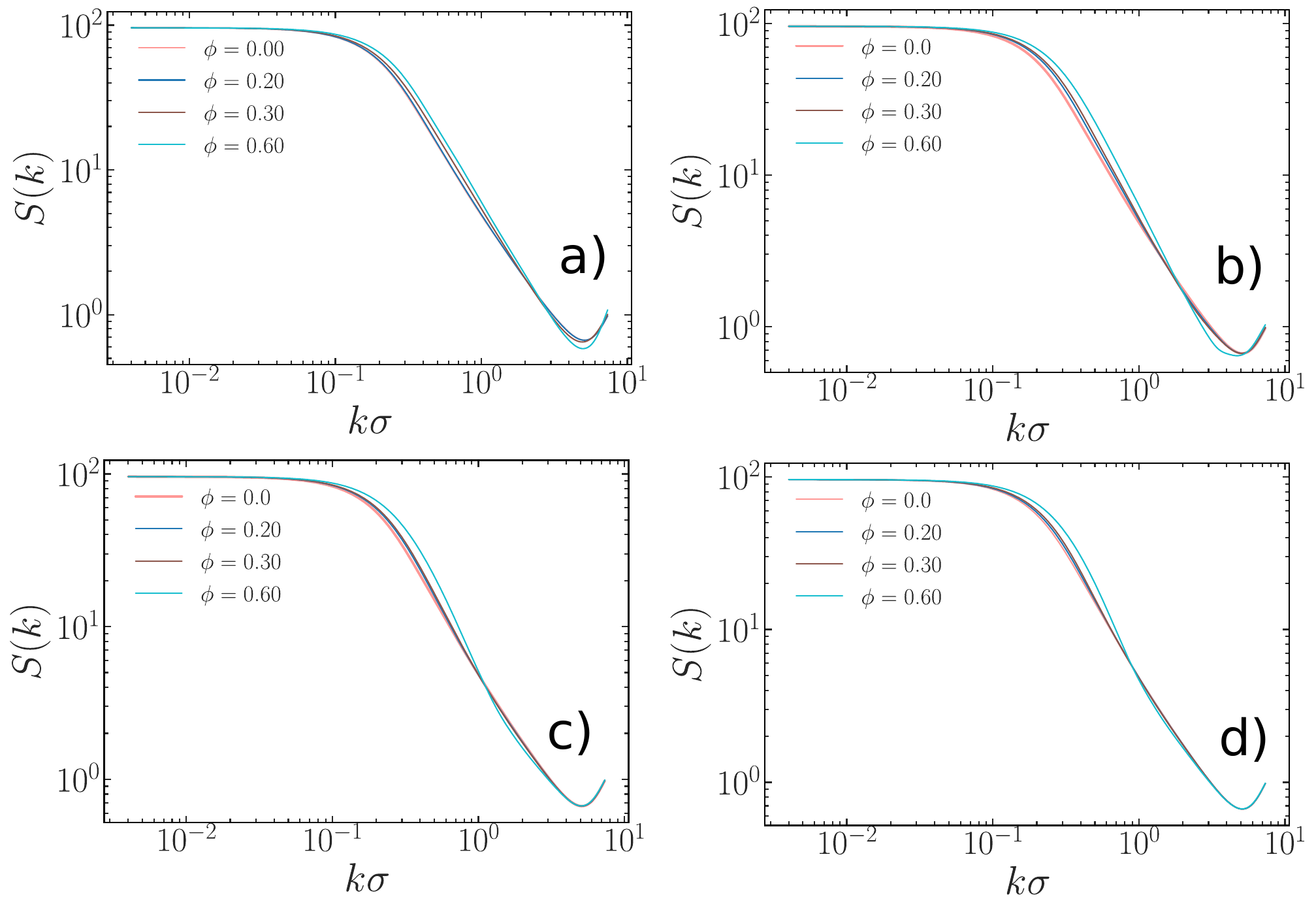}
 		\caption{Plot of \( S(k) \) for \( N = 96 \) at different volume fractions for the Langevin simulations for a) R = 2nm, b) R=4 nm, c) R=12nm , and d) R=18nm.}
 		\label{SK_LG_96}
 \end{figure}
 
 \begin{figure}[tb]

 		\includegraphics[trim= 7 0 0 0,scale=0.255]{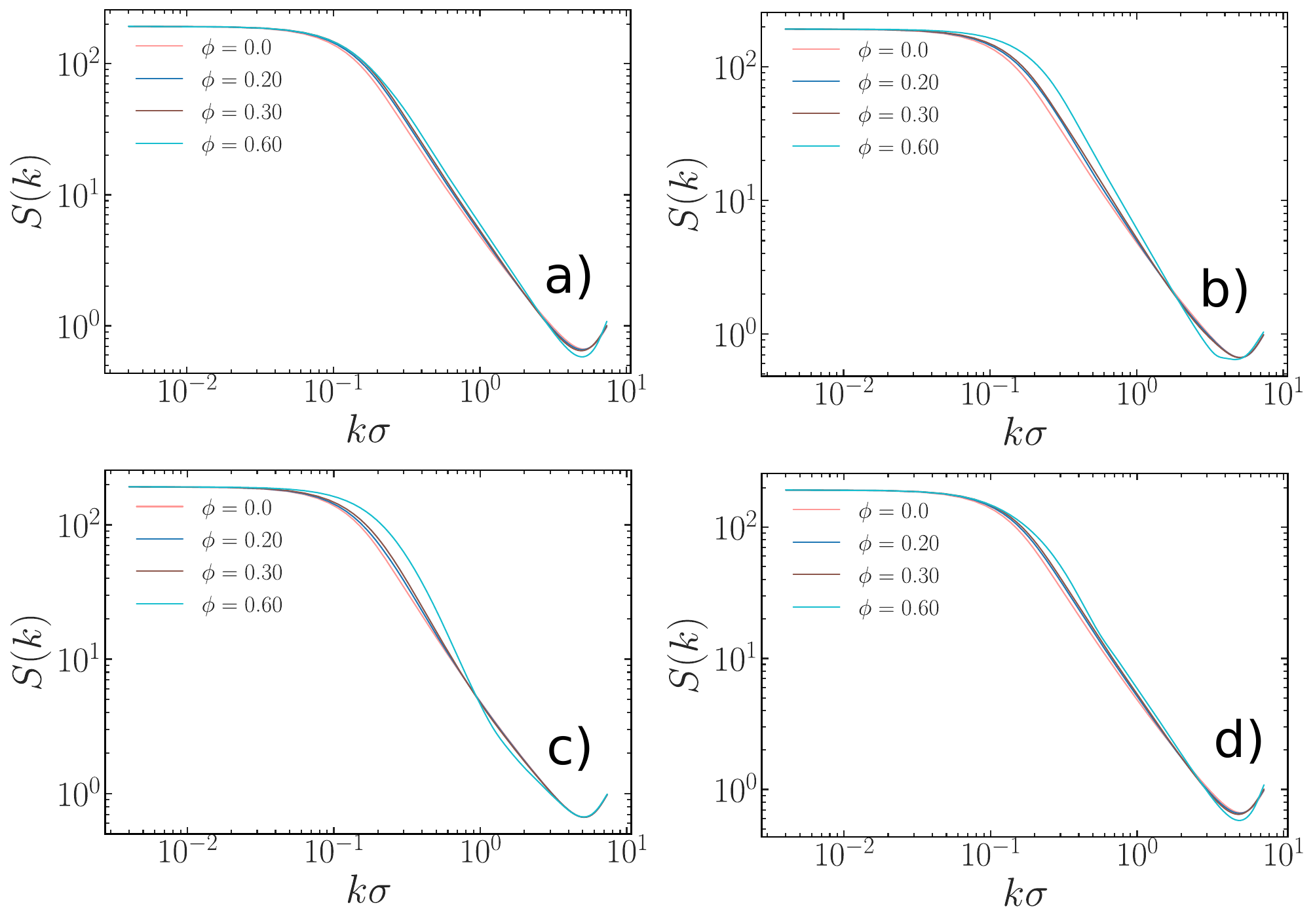}
 		\caption{Plot of \( S(k) \) for \( N = 192 \) at different volume fractions for the Langevin simulations for a) R = 2nm, b) R=4 nm, c) R=12nm , and d) R=18nm.
 		}
 		\label{SK_LG_192}
 \end{figure}
 
\begin{figure}[tb]
 		\includegraphics[trim= 10 0 0 0,scale=0.145]{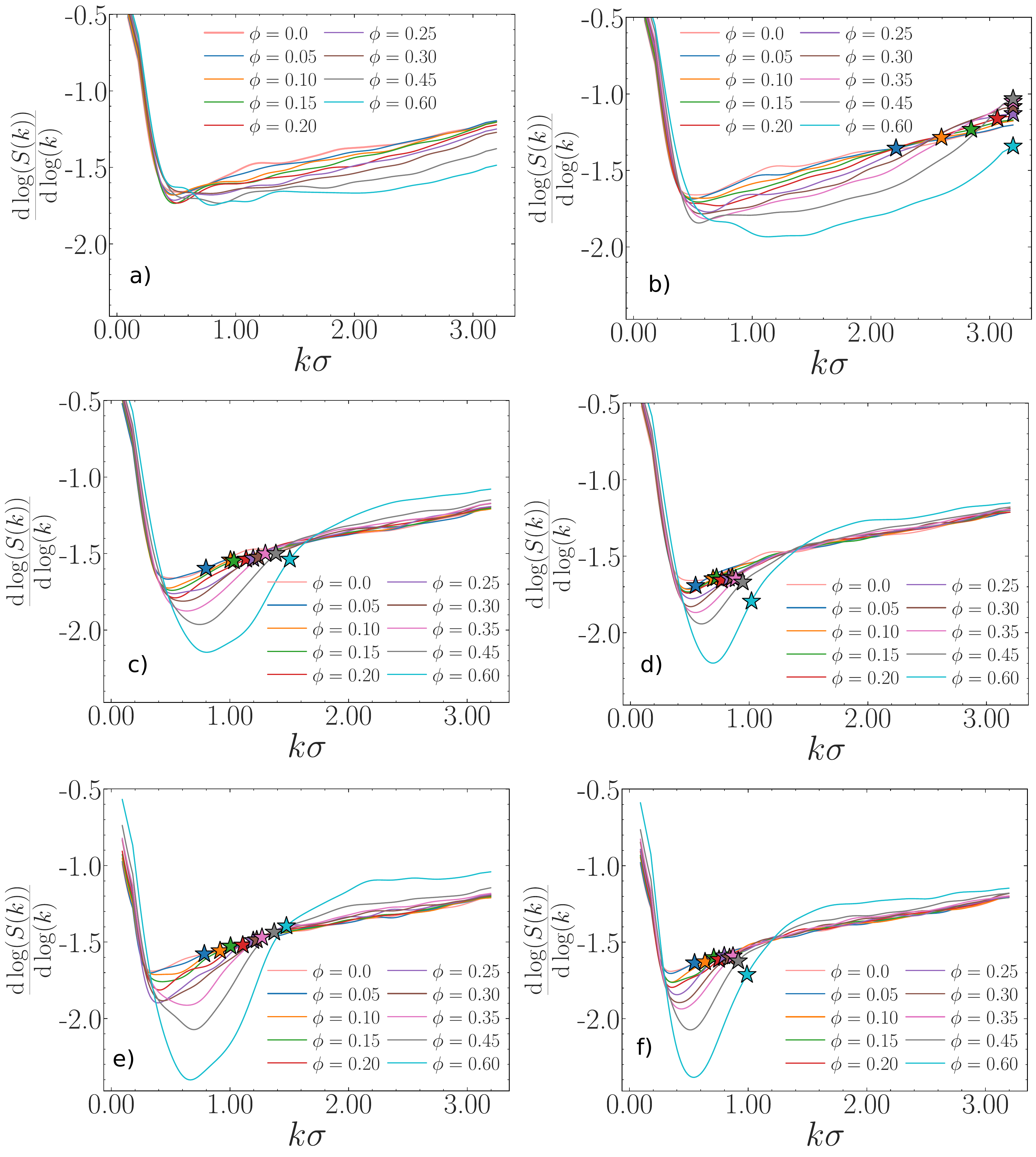}
 		\caption{\( \frac{d \log S(k)}{d \log k} \) for the langevin simulations, expected to be $1/\nu$ in the scaling limit for a chain undergoing a self-similar random walk.  Data for the $N=96$ case and a) $R=2$ nm, b) $R=4$ nm, c) $R=12$ nm, and d) $R=18$ nm.  Data for the $N=192$ case and e) $R=12$ nm and f) $R=18$ nm.  
 		}
 		\label{dlogSK_LG}
 \end{figure}

 Plots of \( S(k) \) for \( N = 96 \) and \( N = 192 \), shown in Figures~\ref{SK_LG_96} and~\ref{SK_LG_192}, illustrate the results for various colloid sizes and volume fractions. There are three regions to these curves.  At low $k$, the plots level off to a plateau.  This small $k$ (long distance) plateau is primarily a finite-size effect due to the short chain lengths.  Consistent with this being a finite-size effect, careful examination shows that the plateau occupies a smaller region (ends at lower values of $k$) for the $N=192$ compared to $N=96$ polymer.   Conversly, the uptick at high $k$ is due to the discreteness of the chains (they are not continuous strings but are made up of discrete monomors).  Between these two extremes there is a fairly linear region (on the log-log scale) of the plot where one can determine the exponent \( -1/\nu \) from \( \frac{d \log S(k)}{d \log k} \) which is plotted in Figure~\ref{dlogSK_LG}.

 \begin{figure*}[ht]

 		\includegraphics[trim= 0 0 0 0, scale=0.20]{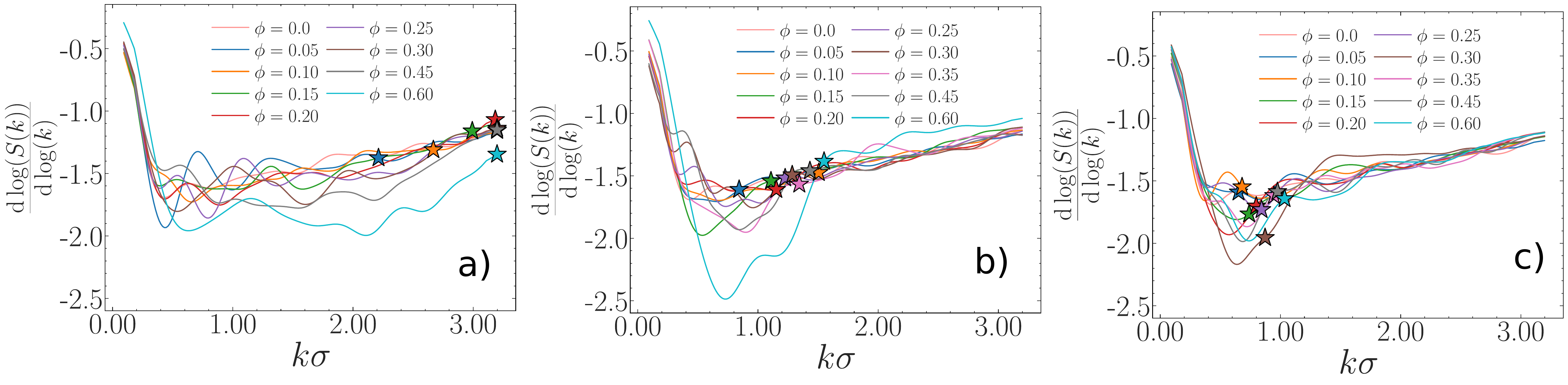}
 		\caption{\( \frac{d \log S(k)}{d \log k} \) for \( N = 96 \) LB simulation at different volume fractions.
 		}
 		\label{dlogSK_LB_96}

 \end{figure*}
 
 \begin{figure}
 	\includegraphics[trim= 20 0 0 0,scale=0.5]{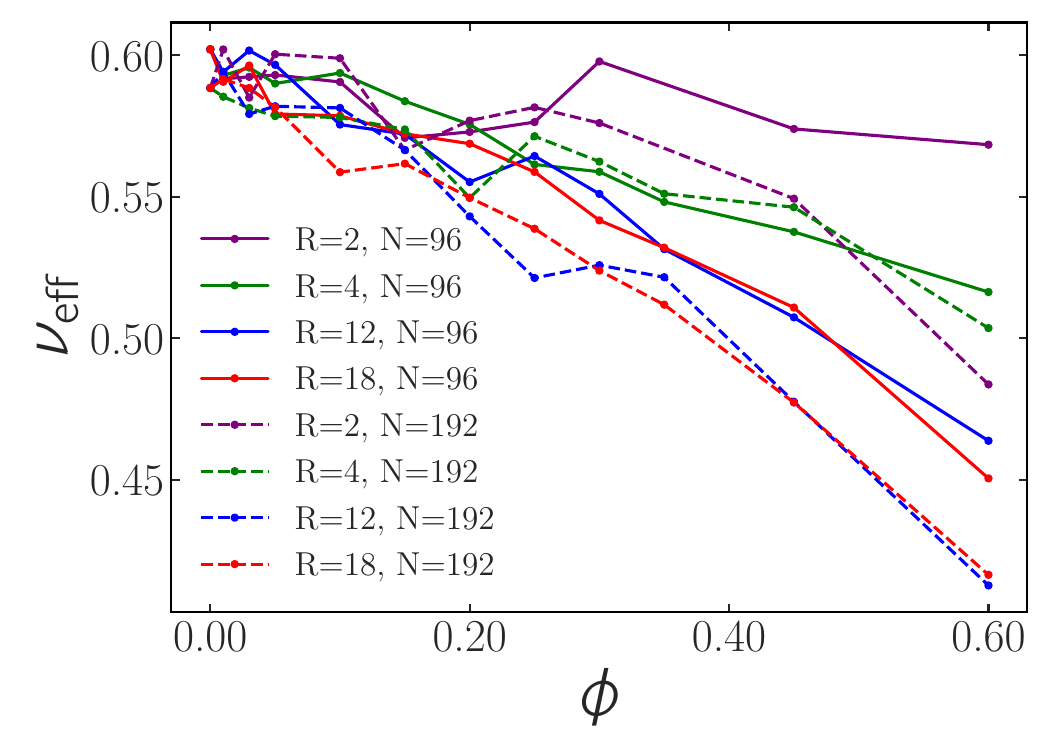}
 	\caption{Effective scaling exponent for the polymer size taken at the smallest $k$ before finite size effects start to dominate.}
 	\label{nueff}
 \end{figure}

At small $k$, the transition from the finite-size plateau to the power law region in the $S(k)$ plot translates in the \( \frac{d \log S(k)}{d \log k} \) plot to a rapid drop from zero to a finite negative value.  This is followed by a gradual linear increase in the mid-to-high \( k \) region.  If the polymer $S(k)$ exhibited the pure power law $k^{-1/\nu}$ scaling over all wavenumbers then this linear region would be flat at a value of \( -1/\nu = -1.7 \approx -1/0.588 \) corresponding to a good solvent.  The behavior seen here at low colloid concentrations are characteristic of semi-flexible chains which are slightly straighter at short wavelengths (so $\nu$ is larger than 0.588) than a true self-avoiding walk (SAW) would display, but do exhibit the characteristic SAW scaling at long wavelengths. At low concentrations, an extrapolation of the gradual linear region would intercept the $k=0$ axis very close to $-1.7$ (i.e. corresponding to $\nu=0.588$) so the expected scaling for a chain in a good solvent is achieved in the long wavelength limit at low colloid concentrations.  
 
At moderate to high concentration of colloids we see distinctly different behaviour for the polymers in the smaller $R=2$ and $R=4$ colloids compared to the larger $R=12$ and $18$ cases.  For the smallest colloids ($R=2$) that are the same size as the monomers that make up the polymer, seen in Fig.~\ref{dlogSK_LG}a, the curves appear to plateau at the asymptotic value of $-1.7$ at a wider range of $k$ as $\phi$ increases (from the smallest values of $k\sigma=0.5$ before finite size effects come into play, all the way up to $k\sigma = 2.25$ for the highest concentration of colloids). This extension of the asymptotic plateau in the d log S(k)/d log k plot toward larger wavevectors indicates that the weak effective local stiffness becomes less pronounced and that self-avoiding-walk-like behavior persists to shorter length scales. For the $R=4$ colloids, we see in Fig.~\ref{dlogSK_LG}b that the effective value of $-1/\nu$ shifts down somewhat across almost the entire range of wavenumbers as $\phi$ increases. Note that a shift down translates to a {\it decrease} in the value of $\nu$ corresponding to more compact configurations.  If we were to interpret the colloidal suspension as a modified "solvent" for the polymer this could be viewed as a degradation in solvent quality from good to something approaching neutral as the concentration of colloids is increased. 

 In contrast, for \( R = 18 \) and \( R = 12 \), the behavior is different: in the intermediate \( k \)-region, the curves shift downward, while at high \( k \), they overlap with the low colloid concentration data. The downward shift for intermediate $k$ (down to values just before finite-size effects kick in) implies that \( \nu < \nu_{\text{free}} \), suggesting that the chain is more compact at the longer wavelengths accessible before finite-size effects kick in, but essentially the same as a free chain at short wavelengths. 
 
 In summary, for $R=2$, the chain compacts primarily at shorter wavelengths.  For \( R = 4 \), the colloids uniformly compact the chain across all wavelengths. However, for \( R = 12 \) and \( R = 18 \), the chain is compacted only at longer (intermediate) wavelengths. For the shorter wavelengths (high $k$) the overlap of the curves at different colloid concentrations is consistent with segments of the polymer acting like free chains in the larger voids between colloids that are present when the colloids are much bigger than the monomer size. 
 
Following the approach used in the LD simulation, Fig.~\ref{dlogSK_LB_96} presents the \( \frac{d \log S(k)}{d \log k} \) plot  at different volume fractions and colloid sizes for LB simulations.   While much noisier than the LD data, the LB data shows the same behavior observed for the LD simulations, as one would expect for a static equilibrium property (only the dynamics are expected to be different for LD and LB simulations.)

To extract the effective scaling exponent at the longest length scales before finite size effects begin to dominate, we take the minimum value of  \( \frac{d \log S(k)}{d \log k} \) as $-1/\nu_{\rm eff}$.  This resulting value of $\nu_{\rm eff}$ is plotted in Figure~\ref{nueff} and is consistent with our observations for the long wavelength behavior of the chain: chains in a solution of smaller colloids, closer in size to a monomer, show only small modifications for the scaling expected for chain in a good solvent ($\nu$ does not drop much from 0.588) while chains in a solution of colloids much larger than the monomer size exhibit scaling suggestive of a worsening of solvent quality with $\nu$ dropping as low as 0.42 in the worst case.  The worsening of the solvent quality is seen to be more significant for the longer chains that can explore longer wavelengths, which also ties in to the enhanced compaction seen in Fig.~\ref{RgvsPhi} for the longer polymers. 

In summary, the compaction of the polymers seen via reduction in $R_g$ as a function of colloid concentration (Fig.~\ref{RgvsPhi}) appears to be mainly a long wavelength effective reduction in solvent quality for the large colloid solutions while for the smaller colloids, the reduction in
$R_g$ is associated primarily with changes in the short-length-scale polymer
conformation, with self-avoiding-walk-like behavior persisting to shorter
length scales.
To further investigate the different behaviours at long and short wavelengths, we next examine the physical length scales present in the colloidal suspension (as opposed to the polymer itself).

\begin{figure*}[tb]
		\includegraphics[trim= 0 0 0 0,scale=0.59]{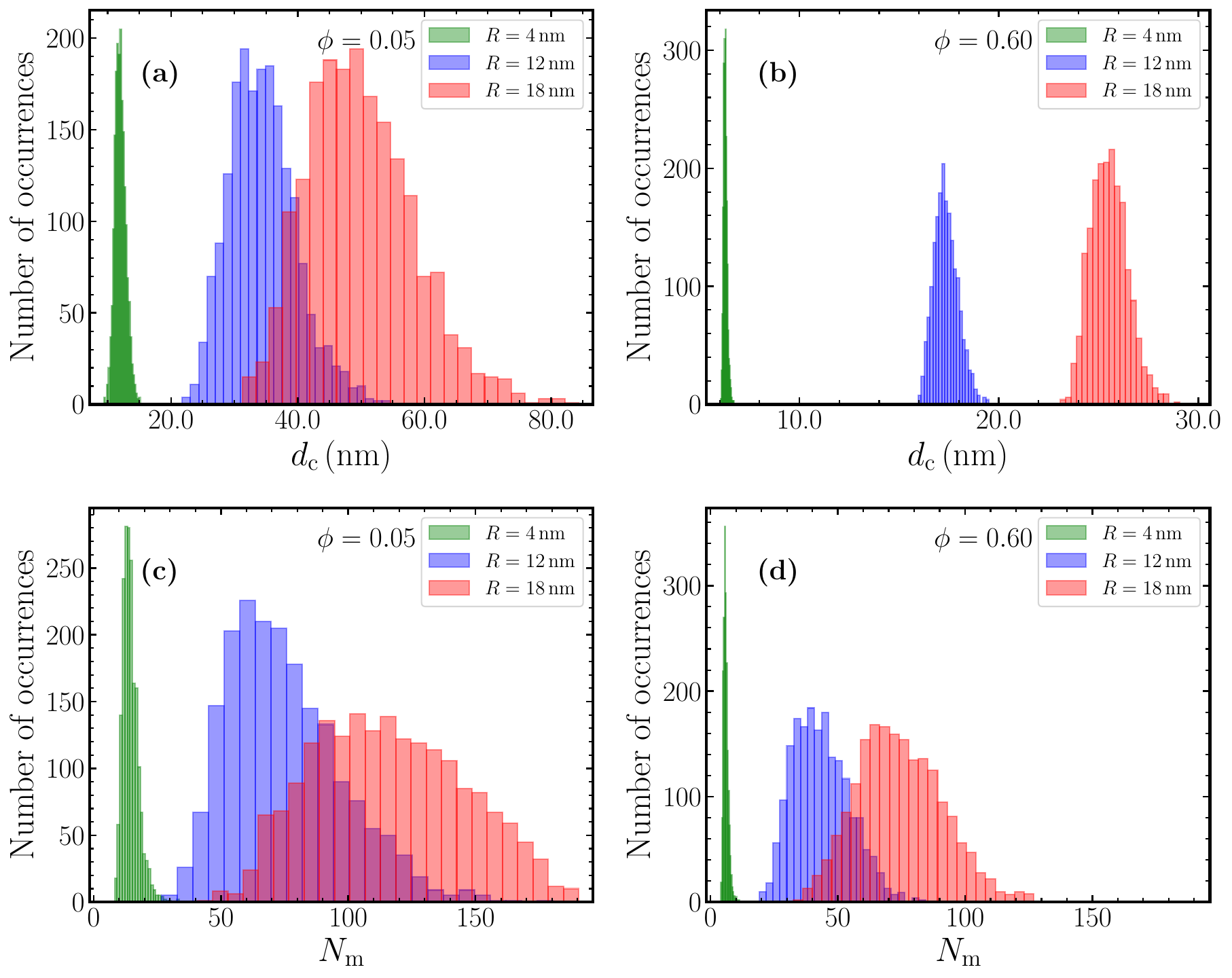}
		\caption{Distributions of the closest-colloid distances, $d_{\mathrm{c}}$, and the corresponding mean number of monomers, $N_{\mathrm{m}}$, for the LD simulations with $N=192$. Here, $d_{\mathrm{c}}$ denotes the closest-colloid distance, and $N_{\mathrm{m}}$ is the mean number of monomers that are closer than the closest-colloid distance. Panels (a) and (b) show the distributions of $d_{\mathrm{c}}$ at crowder volume fractions $\phi=0.05$ and $0.60$, respectively, while panels (c) and (d) show the corresponding distributions of $N_{\mathrm{m}}$. Different colors denote crowder radii $R=4$, 12, and 18~nm.}
		\label{192_d_c}

\end{figure*}

\subsubsection{Length scales in the colloidal suspension} \label{ssec:Nm&dc}

Length scales in the colloidal suspension relevent for the polymer should be connected to the behavior seen in the structure factor $S(k)$ data for the polymer.  In de Gennes blob models of polymers confined in static environments a key concept is that below the length scale of the confinement parts of the polymer contained within "blobs" (of size commensurate with the confinement) behave similar to how they would if the polymer were free \cite{deGennes77, deGennes77b}.  We see signs of this in our examination of the sturcture factor where, particularly for the larger colloids, at short wavelengths the polymer seems fairly unaffected by the colloidal suspension whereas for longer wavelengths it is strongly affected.  To identify with de Gennes blob theory we now examine the relationship between the wavenumber at which we see the crossover of the scaling of the polymers, and the confining length scale of the colloidal suspension. 

In order to estimate this scale, i.e. the blob size of the polymer, we plotted the histogram of the number of monomers that are closer to a given monomer than the closest colloidal particle,and also the distribution of distances from a monomer to the closest colloid, $d_c$. $d_c$ is effectively a measure of the size of the voids, or cages trapping segments of polymer, in the colloidal suspension.  To avoid end effects, the average of the histogram was computed for the central monomers, ranging from monomer 48 to 147 in a polymer with 192 monomers, and from monomer 24 to 71 in a polymer with 96 monomers. The histograms depicted in Figure~\ref{192_d_c} are from a polymer consisting of 192 monomers. The left column corresponds to the lowest crowder volume fraction, $\phi=0.05$, while the right column corresponds to the highest volume fraction, $\phi=0.60$, with both columns showing results for $R=4$, 12, and 18~nm. The upper row shows the distributions of $d_{\mathrm{c}}$, while the lower row shows the corresponding distributions of $N_{\mathrm{m}}$. Histograms for other cases are shown in the supplementary material. Histogams were not plotted for R=2nm as less/no beading along the chain is observed for this case, nor do we see any obvious separation in behavor at different wavenumbers in the structure factor. 

\begin{figure}[tb]
		\includegraphics[trim= 25 0 0 0,scale=0.105]{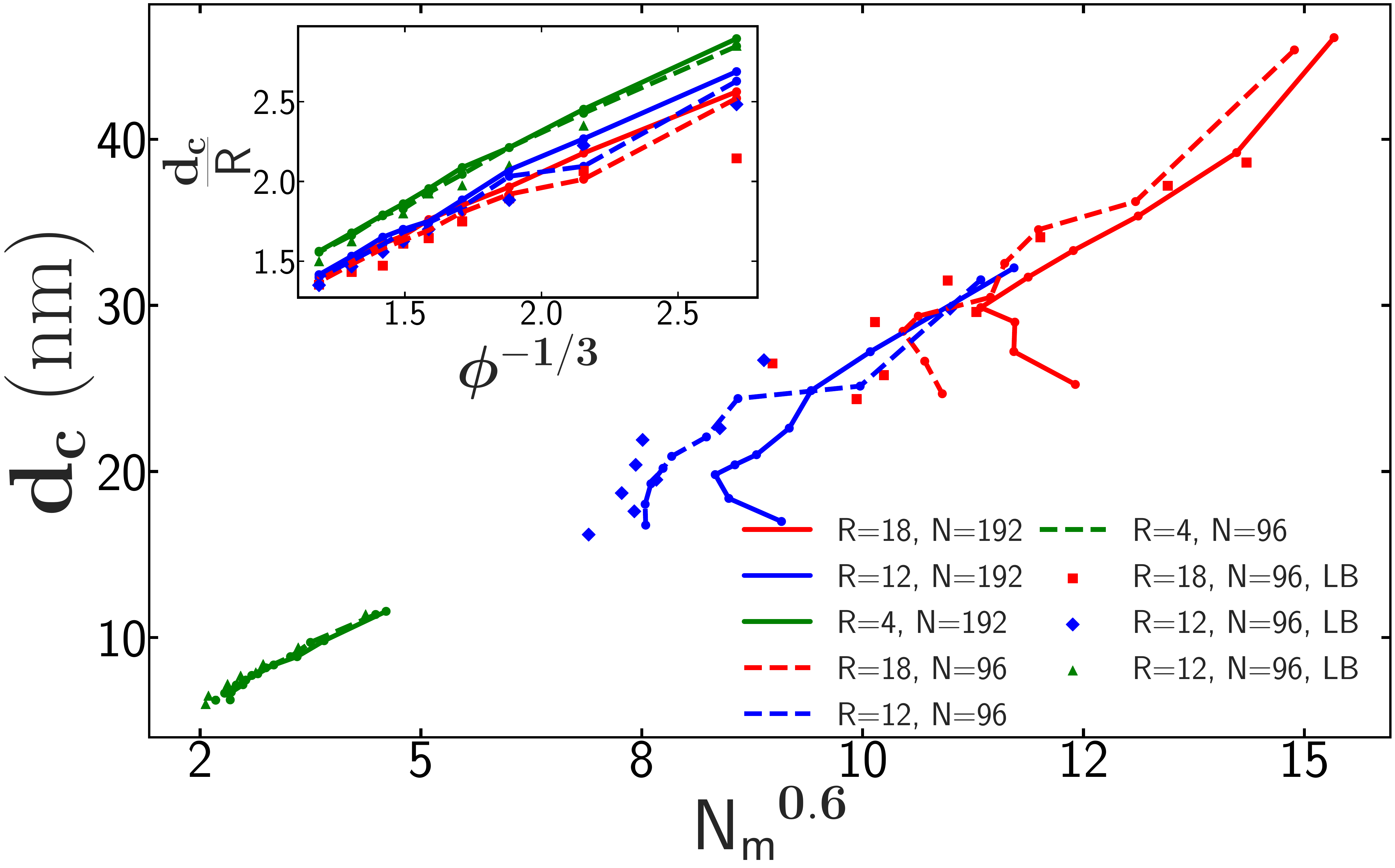}
		\caption{Plot of mean value of closest colloid distance $d_c$ versus number of monomers closer than this distance $N_m^{0.6}$, and (inset) plot of $d_c/R$ versus $\phi^{-1/3}$.
		}
		\label{dc_scale}

\end{figure}

To identify if $d_c$ is indeed the length scale associated with a de Gennes style "blob" we computed the mean of the histograms for different $\phi$ and $R$ and have marked the plots of Figure~\ref{dlogSK_LG} and \ref{dlogSK_LB_96} with symbols at the wavenumber $k=2 \pi/d_c$ corresponding the mean value of $d_c$ for the given $\phi$ and $R$ (note the symbols are colour-coded to match the corresponding curve).  We do see that the symbols do reasonably indicate the positions along these curves where the behaviour meaningfully deviates from the $\phi=0$ chain case giving us confidence that $d_c$ does correspond to a de Gennes "blob" scale.

By comparing the number of monomers located closer than the closest colloid distances, Figure~\ref{192_d_c}, we observe that the range of the histograms decreases as the volume fraction increases for each colloid sizes. A similar trend is evident in the histograms of the closest colloid distances, as shown in Figure~\ref{192_d_c}.
We expect this and $d_c$ to also be related to the distance between colloids, which should be inversely proportional to the cube root of density or volume fraction $\phi$.  This is indeed the case as can be seen in the inset of Fig~\ref{dc_scale} where we see that $d_c/R$ exhibits something close to a linear relationship with $\phi^{-1/3}$. 
To check this further, we plot $d_c$ versus the $N_m^{0.6}$ in Figure ~\ref{dc_scale}, as the number of monomers in a de Gennes "blob" should follow the free polymer scaling \cite{deGennes77, deGennes77b}.  The fairly linear relationship seen in the plot indicates that the blobs do show the expected scaling for the number of monomers in the blob.   

\begin{figure}
    \centering
    \includegraphics[trim= 10 0 0 0,scale=0.44]{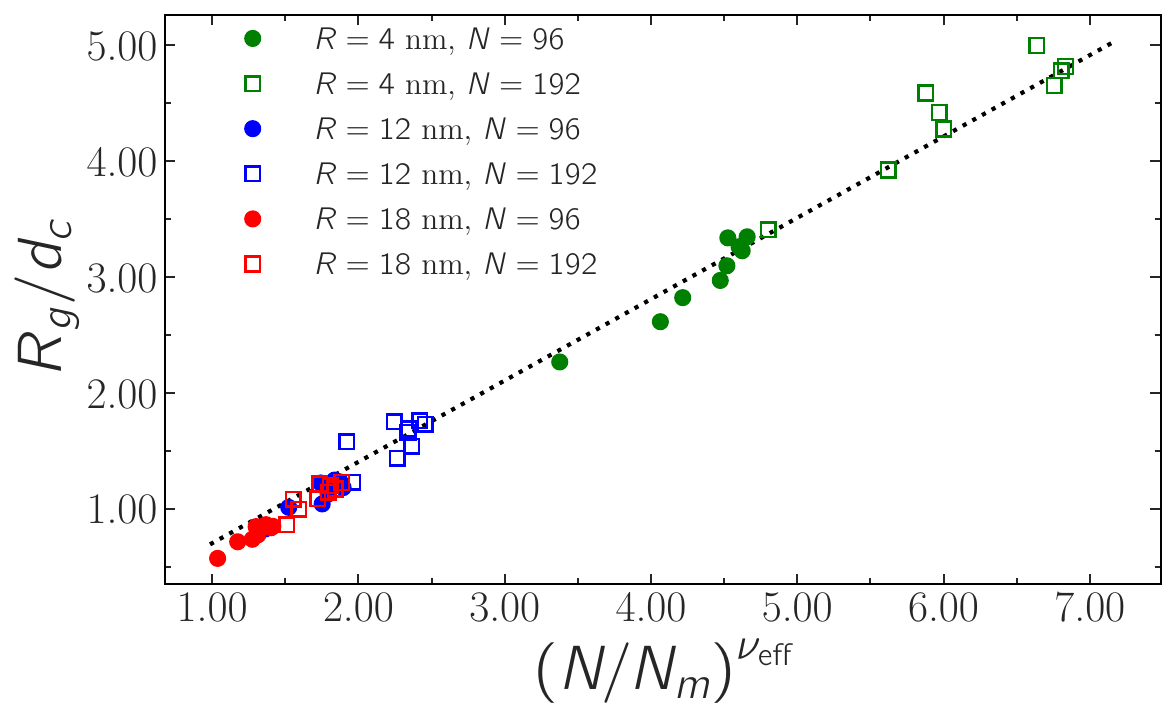}
    \caption{Blob-chain scaling of the polymer radius of gyration LD simulations. The normalized polymer size $R_g/d_c$ is plotted against $(N/N_m)^{\nu_{\mathrm{eff}}}$ for $N=96$ and 192 and crowder radii $R=4$, 12, and 18~nm. The quantities $d_c$ and $N_m$ are obtained from the analysis in Fig.~12, while $\nu_{\mathrm{eff}}$ is obtained independently from the static structure-factor analysis in Fig.~9. The dotted line is a common linear fit through the origin, $R_g/d_c=0.70(N/N_m)^{\nu_{\mathrm{eff}}}$.}
    \label{blob_collapse}
\end{figure}

At long wavelengths, the crowded polymer can be viewed as a chain of de Gennes-style blobs, where each blob has a characteristic size $d_c$ and contains $N_m$ monomers.  This de Gennes blob polymer therefore contains $N/N_m$ effective segments giving the scaling relation
\begin{equation}
    R_g \sim d_c \left( \frac{N}{N_m} \right)^{\nu_{eff}},
    \label{RgScalingBlob}
\end{equation}
analogous to Eq. (\ref{RgScalingPhi0}) for the polymer in bulk solution.  To test this interpretation, we plot $R_g/d_c$ versus $(N/N_m)^{\nu_{eff}}$ in Figure \ref{blob_collapse}, with $\nu_{eff}$ obtained independently from the static $S(k)$ analysis in Fig.~\ref{nueff}.  The results follow a common linear trend that confirms the prediction of Eq.(\ref{RgScalingBlob}).

In summary, the voids in the large colloid suspension form cages that confine sections of the polymer into blobs.  Within these blobs the polymer acts like a free chain.  The larger scale polymer structure is then a modified random walk between these cages.  The random walk of blobs is compacted compared to a simple self-avoiding walk, as indicated by the decrease in the value of $\nu$ seen in the structure factor scaling at long wavelengths. This, in turn suggests that, unlike either the lattice Boltzmann fluid or the virtual solvent of the Langevin simulations that form a "good" solvent for the polymer, the colloid suspension is not a good solvent and the larger colloids are worse solvents than the smaller colloids.

\begin{figure}[tb]
		\includegraphics[trim= 25 0 0 0,scale=0.108]{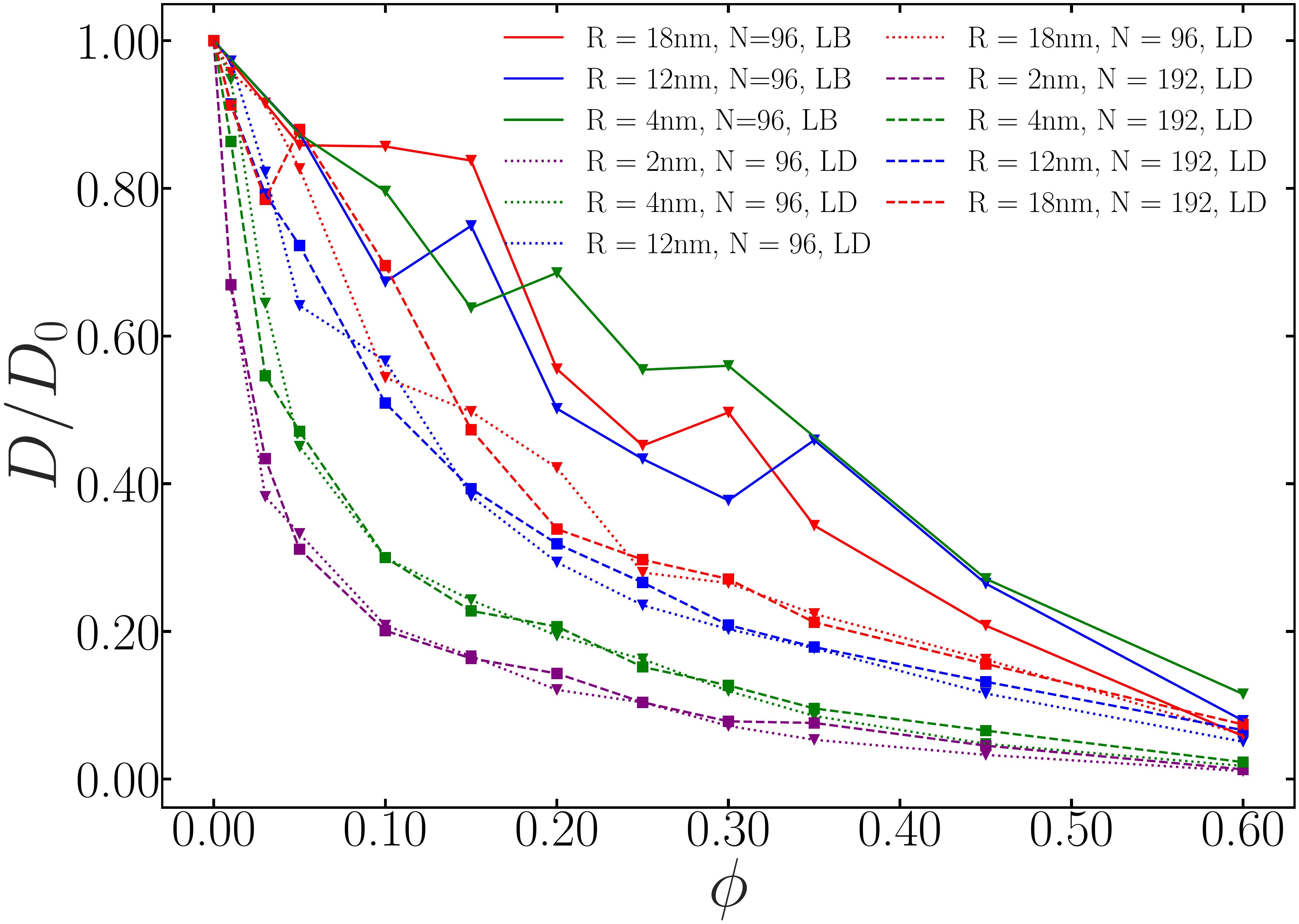}
		\caption{The plot of the diffusion coefficient of the the polymer (D), scaled by the diffusion coefficient of polymer in free solution ($D_0$), as a function of volume fraction of the crowder.
		}
		\label{Dvsphi}

\end{figure}

\begin{figure}[tb]
		\includegraphics[trim= 5 0 0 0,scale=1.05]{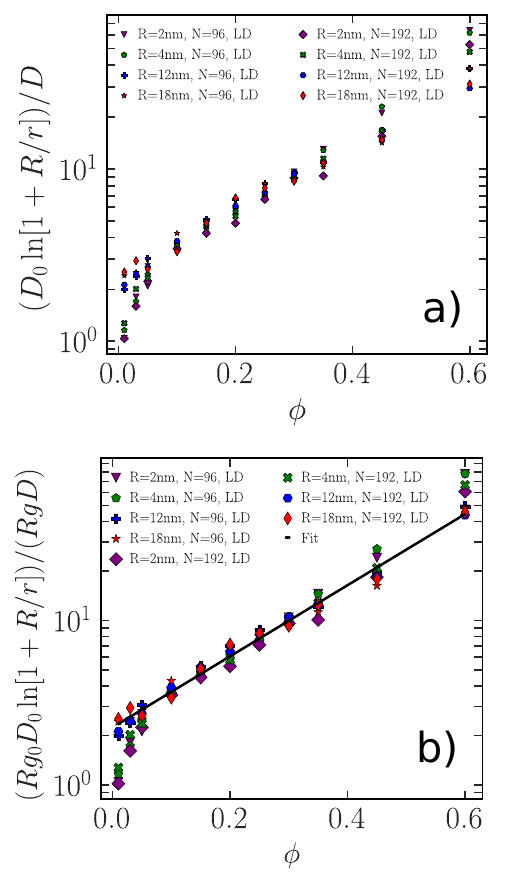}
		\caption{Scaled inverse polymer diffusion coefficient from the Langevin simulations as a function of the crowder volume fraction, $\phi$, for polymer lengths $N=96$ and 192 and colloid radii $R=2$, 4, 12, and 18~nm. (a) The normalized inverse diffusion coefficient, $D_{0}/D$, scaled by the geometric factor $\ln(1+R/r)$, where $D_0$ is the polymer diffusion coefficient in the absence of crowders and $r$ is the monomer radius. (b) The same data with an additional scaling by the polymer radius of gyration, plotted as $R_{g0}D_0\ln(1+R/r)/(R_gD)$, where $R_{g0}$ is the uncrowded radius of gyration. The solid black line shows a common fit to the scaled Langevin data.}
		\label{DScale_LG}
\end{figure}

\begin{figure}[tb]
		\includegraphics[trim= 20 0 0 0,scale=0.52]{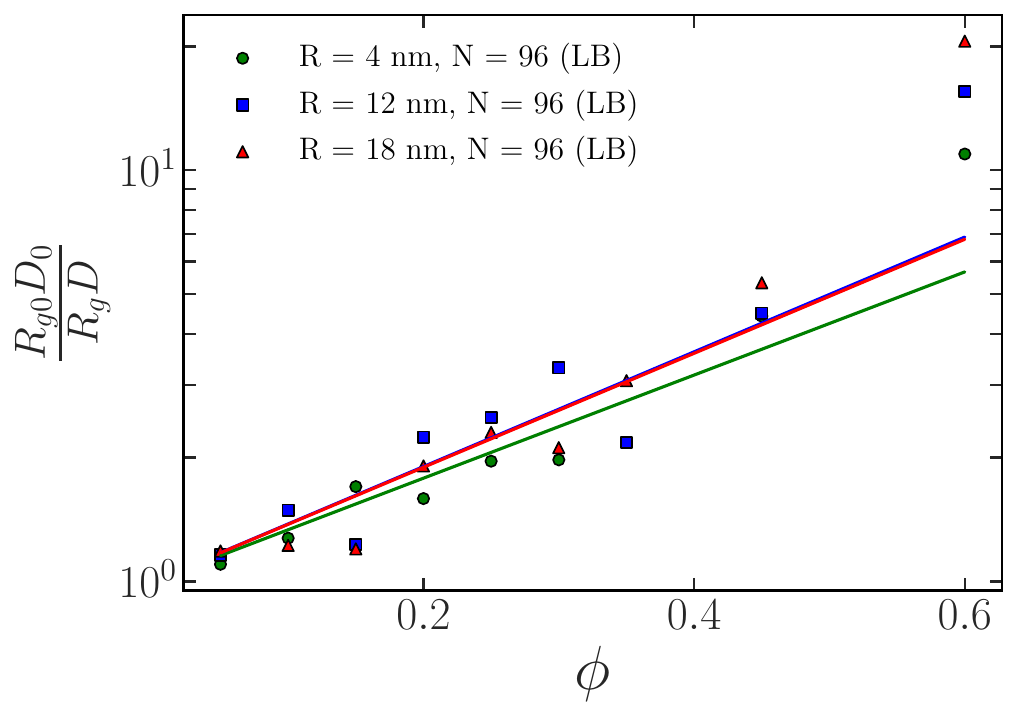}
		\caption{Plot of the inverse of $R_g D$, scaled by the corresponding values in free solution ($D_0$), as a function of volume fraction of the crowder for the lattice Boltzmann data.
		}
		\label{DScale_LB}
\end{figure}

An earlier theoretical study by Zaccone and Terentjev showed that molecular crowding can induce a sharp coil-to-globule transition in a polymer. For Brownian crowders comparable in size to the polymer monomers, their theory predicts that increasing the crowder volume fraction strengthens the effective monomer--monomer attraction, leading to collapse when the effective second virial coefficient changes sign. In their model, this transition occurs at a critical crowder volume fraction of approximately $\phi_c \simeq 0.145$~\cite{Zaccone2012}. Kang et al.~\cite{Kang2015} also showed that crowding can induce a coil-to-globule transition and introduced the dimensionless variable $x = R_{g0}/d$, where $d$ is the crowder spacing.
Kang et al.~\cite{Kang2015} also investigated how macromolecular crowding affects polymer conformation and showed that crowding can drive a coil-to-globule transition. They defined a dimensionless variable $x=R_{g0}/d$, where $d$ is the crowder spacing ($d \sim d_c$ in our notation).  For the systems we examine here the largest value of $x$ is $10.5$, which remains well below the value $x_c=17$ at which Kang et al. observed a cooperative collapse.   Most of our systems therefore lie in the gradual-compaction regime rather than the stronger collapse regime emphasized in their work. This also explains why the compaction curves for the different crowder radii remain relatively close to one another in Fig.~\ref{RgvsPhi}. Thus, the more modest compaction observed here is consistent with the different range of size ratios and crowding conditions explored in our simulations.

This analysis also helps explain why the scaling proposed by Kang et al. does not fully collapse our data. Their scaling incorporates the changing inter-crowder separation $d$, which corresponds to our $d_c$, but assumes that the conformational exponent $\nu$ remains unchanged under crowding. In contrast, our static $S(k)$ analysis shows that $\nu_{\mathrm{eff}}$ varies with the crowder conditions. Our results, completely consistent with de Gennes blob theory, therefore extend the scaling picture of Kang et al. by showing that both the colloidal spacing and the changing long-wavelength polymer exponent must be included to describe the conformational response.
 
\subsection{Dynamics}
\subsubsection{Diffusion}

Figure~\ref{Dvsphi} shows a plot illustrating the diffusion coefficient of the center of mass of the polymer normalized by $D_0$ (the diffusion coefficient of the polymer in a free solution), as a function of the volume fraction of crowders. Unlike static equilibruim data on the polymer and colloidal configurations, the very different dynamics for the lattice Boltzmann (which includes hydrodynamics) and Langevin simulations means there is no expectation for the diffusion constant to behave similarly for these systems.  Interestingly, diffusion in the Langevin simulations drops at a much more rapid rate at low colloidal volume fraction compared to diffusion in the Lattice Boltzmann simulations.  Further, diffusion in the Langevin simulations shows a very strong dependence not just on $\phi$, but also on the size of the individual colloids $R$, a dependence that is not observed in the LB data.  In particular for small colloids, the Langevin dynamics show a rather exteme drop in diffusion at low $\phi$.

If we were to assume the polymer diffusion in the Langevin simulations followed the same scaling as in a solution without colloids, Eq. (\ref{DLG}), but just with a higher drag coefficient $\gamma_\phi$ then we would expect
\begin{equation}
	D_0/D = \gamma_\phi/\gamma_0.
	\label{gamma_phi}
\end{equation}
Unlike the Langevin case, polymer diffusion in lattice Boltzmann also includes hydrodynamics and depends on $R_g$ in Eq.(\ref{DLB}).  If we assume that the scaling remains the same but the effective viscosity $\eta_\phi$ changes as a function of colloid concentration then we would expect
\begin{equation}
	D_0 R_{g0}/(D R_g) = \eta_\phi/\eta_0.
	\label{DRg}
\end{equation}  
Low molecular weight polymer self-diffusion (polymer diffusing in a suspension of similar polymers)  tends to follow an exponential dependance on the concentration of polymers \cite{Phillies89,Phillies98}. Effective medium arguments \cite{Adler80} to explain this dependance, that would not be very different if the polymer were in a suspension of colloids of similar size rather than similar polymers, would also indicate the effective visocity experienced by the polymer might be expected to follow
\begin{equation}
	\eta_\phi/\eta_0 \sim \exp(b \phi),
	\label{etaphi}
\end{equation} 
for some constant $b$.  

Based on the expected exponential dependance on $\phi$ we plot the inverse of the scaled diffusion constant in Fig.~\ref{DScale_LG} for the Langevin simulations and in Fig.~\ref{DScale_LB} for the lattice-Boltzmann simulations.  The data for the lattice Boltzmann simulations in Fig.~\ref{DScale_LB} appear to reasonably follow such a relationship, except at $\phi=0.6$ (we will examine that case below).  

Unlike the LB case, langevin dynamics results in a strong dependence on the size of the crowders.  We have added the scaling factor of $\ln(1 + R/r)$ for the Langevin data as, phenomenologically, this appears to provide a reasonable collapse for different sized colloids.  Based on Fig~\ref{DScale_LG}a and Eq.(\ref{gamma_phi}), this suggests that for the Langevin simulations the data can be reasonably fit to
\begin{equation}
	\gamma_\phi \approx \frac{\gamma_0}{\ln (1+R/r)} \exp(-b_\gamma \phi),
\end{equation} 
with the exception of the low volume fraction data for the smallest colloids.  There is some scatter at higher $\phi$ in Fig~\ref{DScale_LG}a.  To investigate if there could be some dependence on polymer size appearing  we plot in Fig~\ref{DScale_LG}b the langevin data scaled by $R_g$, similar to how we scaled the LB data.  This does appear to reduce the scatter somewhat but it is difficult to determine if this is statistically significant.  Such a depenance on $R_g$ could be justified by an effective hydrodynamic-like interaction induced over the scale of the polymer via the presense of colloids.  Indeed, colloids cannot overlap and, as a result, movement of one colloid will induce correlated motion in a colloid in close proximity.  This might also explain the fact that diffusion in the small colloids at very low volume fraction ($\phi \lesssim 0.05$) does not fit the same pattern as the rest of the data.  Small colloids at low density are too far apart to directly interact and would not exhibit any correlated motion with Langevin dynamics.  In contrast, the larger colloids are already of a comparable size to the radius of gyration of the polymer and hence any close interaction of a polymer with even one colloid will induce some correlated motion in monomers.  At moderate to higher densities, all the colloidal suspensions contain regions of closely interacting colloids and could demonstrate hydrodynamic modes at the length scales experienced by the polymer.  We will examine this further in the next subsection.

\begin{figure}[ht]
		\includegraphics[trim= 0 0 0 0,scale=0.5]{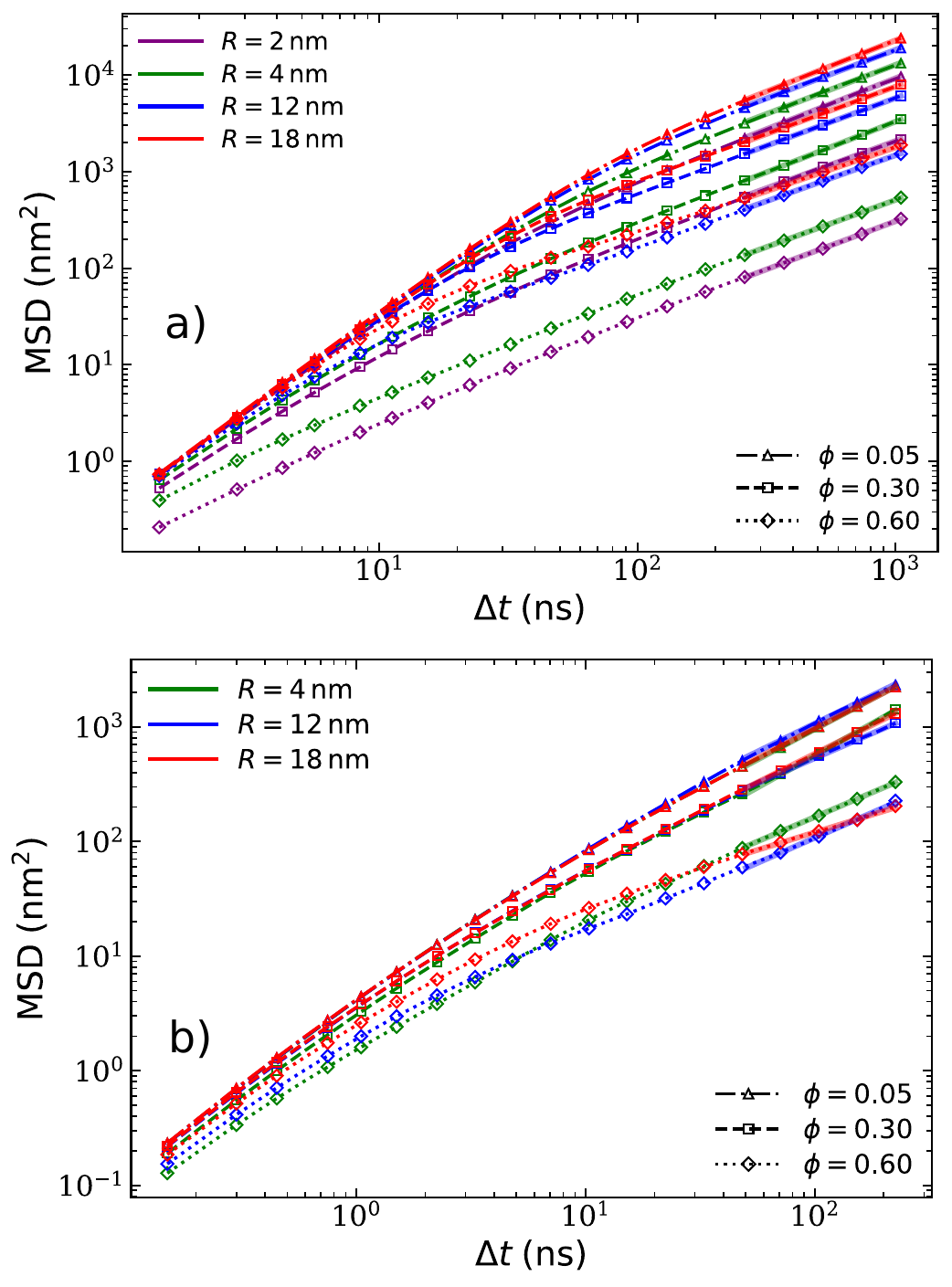}
		\caption{Mean-squared displacement (MSD) of the polymer center of mass as a function of lag time $\Delta t$ on a log--log scale for selected crowder volume fractions $\phi=0.05$, 0.30, and 0.60. (a) Langevin-dynamics (LD) simulations for $N=192$ with crowder radii $R=2$, 4, 12, and 18~nm. (b) Lattice-Boltzmann (LB) simulations for $N=96$ with $R=4$, 12, and 18~nm. Colors indicate the crowder radius, while line styles and symbols distinguish the different volume fractions. The long-time MSD is fitted to $\mathrm{MSD}\sim(\Delta t)^\alpha + C$ to determine the diffusive scaling exponent $\alpha$.}
		\label{MSD_polymer}
\end{figure}

Before looking at the dynamic scaling, we note that the polymer diffusion at $\phi=0.6$ is slower than you would expect based on the trends at other volume fractions (the points for the inverse of the scaled diffusion constant in Fig.~\ref{DScale_LG} and \ref{DScale_LB} are above the trend line). To check that the polymer motion is not experiencing sub-diffusive behavior, the mean-squared displacement (MSD) versus time is shown in Figure~\ref{MSD_polymer} for the LD (a) and LB (b) simulations.  The only case where the polymer motion is significantly sub-diffusive at long times is the LB simulation at $\phi=0.6$ with $R=18 nm$.  In contrast, we looked at the MSD for individual colloids and their motion remains diffusive for all cases (see supplementary material) so this behavior is related to the polymer itself and hydrodynamic effects appear to play a role (as the LD simulation for the same case remains diffusive).  First, we note that $\phi=0.6$ is approaching random-close packing ($\phi \sim 0.63$) and for the $R=18 nm$ colloids the cages are big enough to fit almost the entire polymer chain, i.e. $d_c \sim R_g$. As a result, the chain is effectively a single de Gennes-style blob, about the same size as a cavity, and moves more like a free chain inside a single cage.  To escape the cage part of the chain must move through narrow regions between cages.  In the LD simulations these regions do not offer more resistance to monomer motion than any other region.  However, in the LB simulations the flow in these regions is restricted by the close proximity to neighboring colloid surfaces (the same effect that gives rise to lubrication forces) which is likely to make escape from a cage take longer.
One might expect chain reptation to be the only remaining mode for moving the chain.  However, watching a movie of this simulation (Figure~\ref{LB_pol} (Multimedia available online)), we notice that loops between beaded sections (i.e. clusters of monomers in nearby cages) sometimes slide over large colloids in a direction perpendicular to the chain. This movement isn’t really reptation as it involves transverse loops sliding normal to the tangent along their length.  These loops can then nucleate a new beaded section of monomers in a new cage.  In any case, it appears that the polymer dynamics are entering a new regime as the colloids approach the immobility associated with jamming. 

\begin{figure}[ht]
		\includegraphics[trim= 10 0 0 0,scale=0.95]{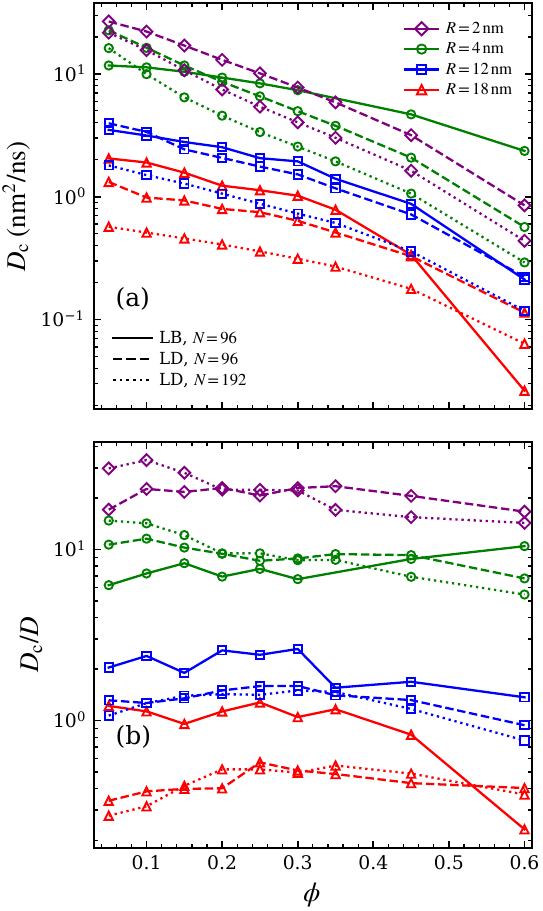}
		\caption{(a) Colloid diffusion coefficient, $D_{\mathrm{c}}$, as a function of crowder volume fraction, $\phi$, on a semi-logarithmic scale. (b) Ratio of the colloid diffusion coefficient to the corresponding polymer diffusion coefficient, $D_{\mathrm{c}}/D$, as a function of $\phi$. Different colors and symbols denote the crowder radius $R$, while line styles distinguish the LB and LD simulations and polymer lengths.
}
		\label{D_colloids}
\end{figure}

To better understand the polymer diffusive behavior, we directly compare the mobilities of the crowders, shown in Figure~\ref{D_colloids}a, and the polymer. Figure~\ref{D_colloids}b shows the ratio $D_c/D$, where $D_c$ is the crowder diffusion coefficient and $D$ is the polymer center-of-mass diffusion coefficient at the same $\phi$. The horizontal line at $D_c/D=1$ separates systems in which the crowders diffuse faster than the polymer from those in which they diffuse more slowly. For the smaller crowders, $R=2$ and $4\,\mathrm{nm}$, the ratio is substantially greater than unity, indicating that the crowders rearrange considerably faster than the polymer. For $R=12\,\mathrm{nm}$, $D_c$ and $D$ are generally comparable. For $R=18\,\mathrm{nm}$, $D_c/D$ is often below unity, particularly at large volume fractions, showing that the largest crowders become strongly constrained and can diffuse more slowly than the polymer.
The relative crowder mobility therefore decreases with increasing crowder radius and, for the largest crowders, with increasing volume fraction. Small, rapidly diffusing crowders continually reorganize around the polymer, whereas slowly diffusing large crowders generate a more slowly relaxing local environment. At high volume fractions, the motion of the largest crowders is restricted by geometric confinement and caging. In the LB simulations, hydrodynamic lubrication resistance may provide an additional contribution when neighboring colloid surfaces approach closely, because the fluid confined within the narrow gap must be displaced as the particles move relative to one another. This near-contact hydrodynamic resistance slows the relative motion of the colloids. Such lubrication effects are absent in the LD simulations, where the solvent is implicit and hydrodynamic interactions are not included. The reduction in crowder mobility accompanies the stronger suppression of polymer center-of-mass diffusion observed for large $R$ and high $\phi$.

\subsubsection{Polymer dynamics scaling}

To better understand the dynamics as a function of wavelength, a standard Rouse mode analysis for the chain has been implemented.  Rouse modes, the normal modes of oscillation along the chain, are defined via~\cite{Kopf1997}:
\begin{equation}
	X_p = \frac{1}{N} \sum_{m=1}^{N} r_m \cos \left( \frac{p\pi}{N} \left( m - \frac{1}{2} \right) \right) , 
\end{equation}
where $r_m$ is the position of the $m^{th}$ monomer and \(p = 1, 2, \cdots, N - 1\) is the mode number. For the Zimm model, these modes are expected to decay exponentially with a relaxation time \( \tau_p\) as follows:
\begin{equation}
	\langle X_p(t+s) \cdot X_p(s) \rangle = \langle X_p^2 \rangle e^{-t/\tau_p}
\end{equation}

\begin{figure*}[ht]%[ht]
	%	\begin{center}
		\includegraphics[trim= 100 0 0 0,scale=0.11]{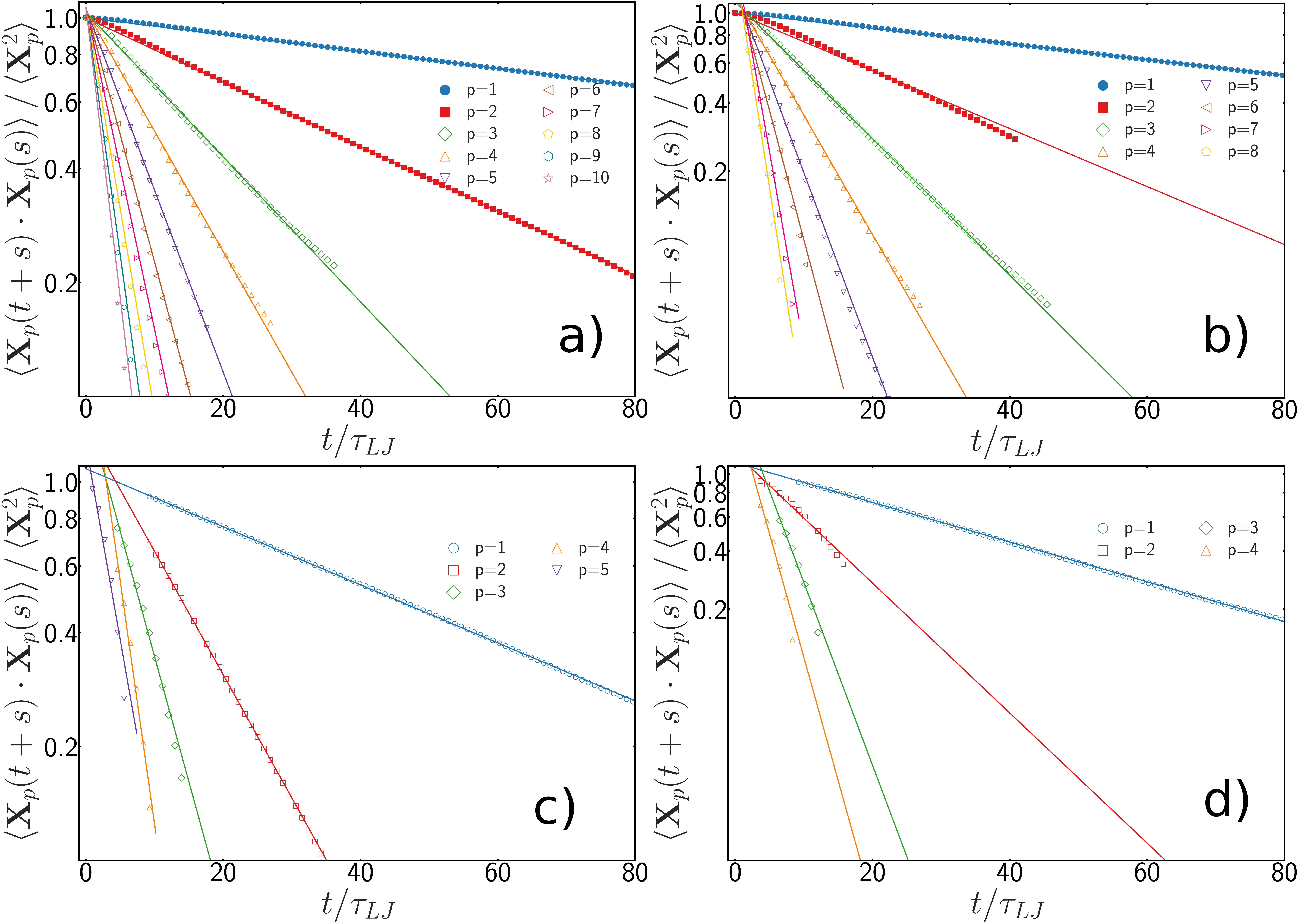}
		\caption{Rouse mode correlation functions for modes $p = 1$ to $10$ from Langevin dynamics simulations at a volume fraction $\phi = 0.20$ for N=96. The panels show data for different colloid sizes in  in a log-linear plot: a) $R = 2$, b) $R = 4$, c) $R = 12$, and d) $R = 18$.}
		\label{ROUSE_LG_96}
		%	\end{center}
\end{figure*}

\begin{figure*}[ht]
		\includegraphics[trim= 110 0 0 0,scale=0.05]{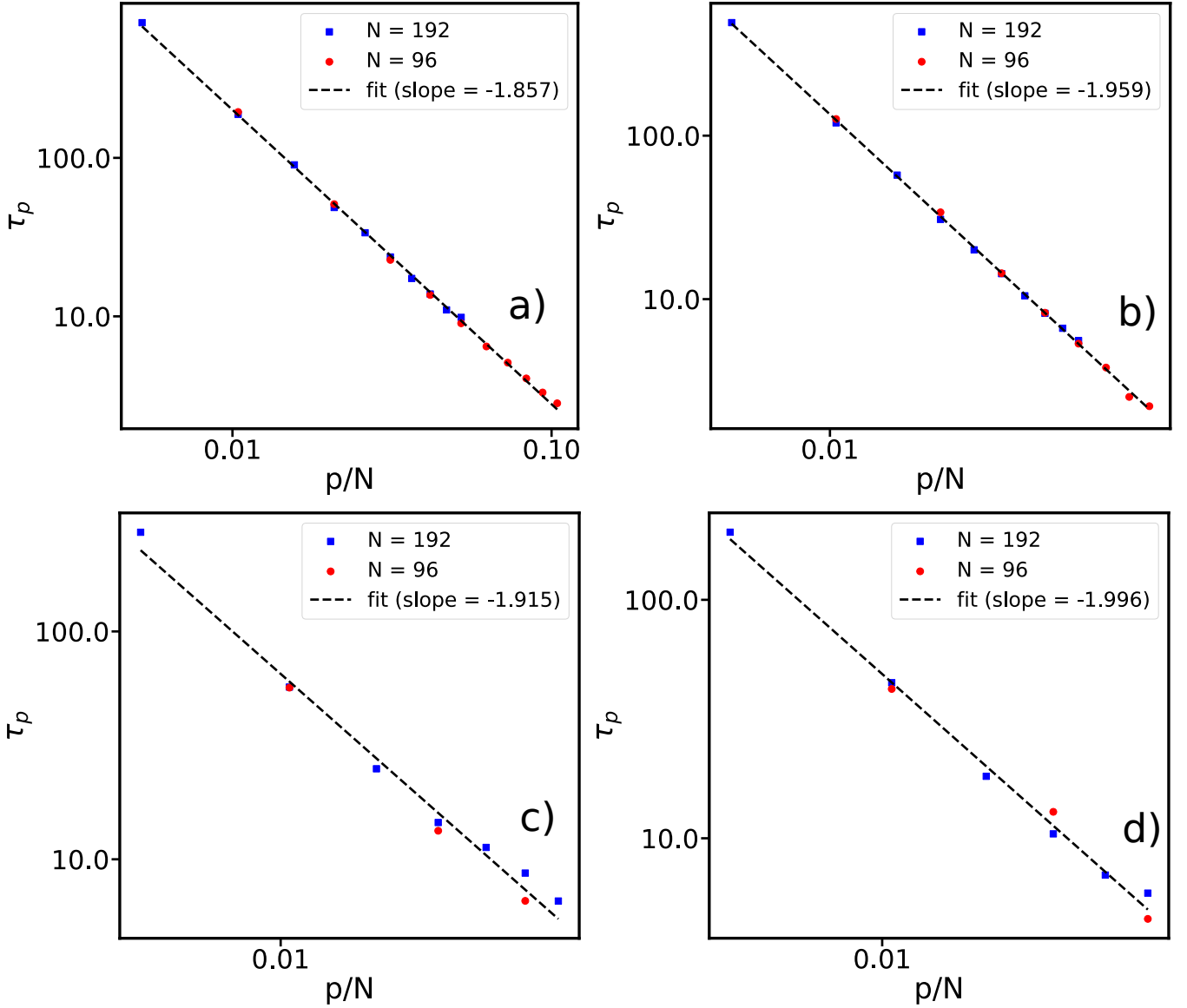}
		\caption{Log-log plot of the Rouse mode relaxation time, $\tau_p$, as a function of the normalized mode number, $p/N$. Data is from Langevin dynamics simulations for polymer length $N=96$ and $N=192$ at a volume fraction $\phi = 0.20$, shown for different colloid sizes: a) $R = 2$, b) $R = 4$, c) $R = 12$, and d) $R = 18$. 
		}
		\label{tau_p_LG_96}

\end{figure*}

In the Zimm model, thermal fluctuations are introduced through noise acting directly on monomers, leading to independent relaxation of different polymer modes and resulting in a diagonal mode-coupling matrix. In contrast, the fluctuating lattice Boltzmann (LB) model does not apply explicit Langevin noise to the monomers; instead, fluctuations arise from the fluid itself, leading to a non-diagonal mode-coupling matrix. As a result, the LB model (or any fully hydrodynamic model) does not create a one-to-one mapping to Zimm theory. Despite this, it is useful to examine how the measured relaxation times align with theoretical expectations for scaling~\cite{Polson2006}:
\begin{equation}
	\tau_p^{-1} \sim (p/N)^{z \nu} r_{\nu}(p), 
	\label{tau_scaling}
\end{equation}
where $z$ is the dynamic scaling exponent that depends on
chain rigidity, \( \nu \) is again the Flory exponent, and \( r_\nu (p)\)
a slowly varying function of $p$.  This analysis has been done previously for the polymer in a bulk LB fluid \cite{Ollila2011-wg} where the value of $z=3$ was found, consistent with the Zimm model where hydrodynamics are included.  Rouse dynamics are expected for a polymer undergoing Langevin dynamics where $z=3.7$ \cite{Doi}.

We computed the Rouse mode correlation functions for $p=1-10$ and plotted them in a log-linear scale for polymer in bulk as well as polymer surrounded with the various colloid sizes at different overall volume fractions. The relaxation times were measured by fitting the data linearly on a log-linear plot. As an example of the anaylsis, Figures ~\ref{ROUSE_LG_96} and ~\ref{tau_p_LG_96} show the data for Langevin dynamics with $\phi = 0.20$ for N=96, and N=192 with $\phi = 0.20$.  The data for other cases are shown in the supplementary material.

A detailed analysis of the polymer's Rouse modes for the Langevin case reveals a significant dependency on the colloid volume fraction ($\phi$) and colloid size. At constant colloid size but sufficiently high volume fractions, the Rouse mode correlation functions for all modes are observed to follow the expected scaling relationship.  However, at low volume fractions ($\phi$), we observe a breakdown of this scaling behavior. This breakdown is not uniform across all modes; it is most pronounced for short-wavelength modes (high $p$), while long-wavelength modes (low $p$) remain more consistent with the scaling law. Also, the scaling behavior holds for smaller colloid sizes rather than large ones. We can attribute this phenomenon to the sparse nature of polymer--colloid interactions at low $\phi$. At low $\phi$ and large colloid sizes, any sub-chain shorter than a critical length does not interact with colloids frequently enough for a clear scaling to hold. In contrast, for a given colloid size at high enough $\phi$ the functions show a clear linear region (in a log-linear plot) beginning near $t = 0$. At low $\phi$ and large colloids, this linear region only appears at longer lag times. Consequently, our fitting procedures were adapted: data with clear scaling over all times (closed markers) were fit with the constraint that the Rouse mode correlation approaches 1 at the origin, while data that showed scaling only at longer time scales (open markers) were fit only in the linear region, relaxing this constraint. In Figures ~\ref{ROUSE_LG_96}, the solid markers correspond to modes whose correlations are linear from t=0 and are fit with the constraint that the Rouse mode correlation function equals 1 at the origin. The open markers represent modes that become linear only at longer time scales and are fit without imposing this constraint, as explained earlier.  Data points at high $p$ with no linear region were excluded from the final scaling analysis.

In contrast, the expected scaling behavior for the polymer in the Lattice Boltzmann (LB) fluid is found to hold for all mode numbers $p$ and across all volume fractions and for polymer in bulk (see the supplementary material for the plots). The reason for this behavior probably lies in the inclusion of hydrodynamic interactions (HI), which are intrinsic to the LB method. The fluctuating fluid itself mediates a long-range coupling between all monomers and colloids (at any $\phi$), resulting in a non-diagonal mode-coupling matrix. This means the polymer modes are not independent but are coupled by the solvent. This fluid coupling then ensures that the scaling behavior is maintained for all modes, even at low $\phi$ where direct colloid interactions would be sparse.

\begin{figure}[ht]
		\includegraphics[trim= 100 0 0 0,scale=0.15]{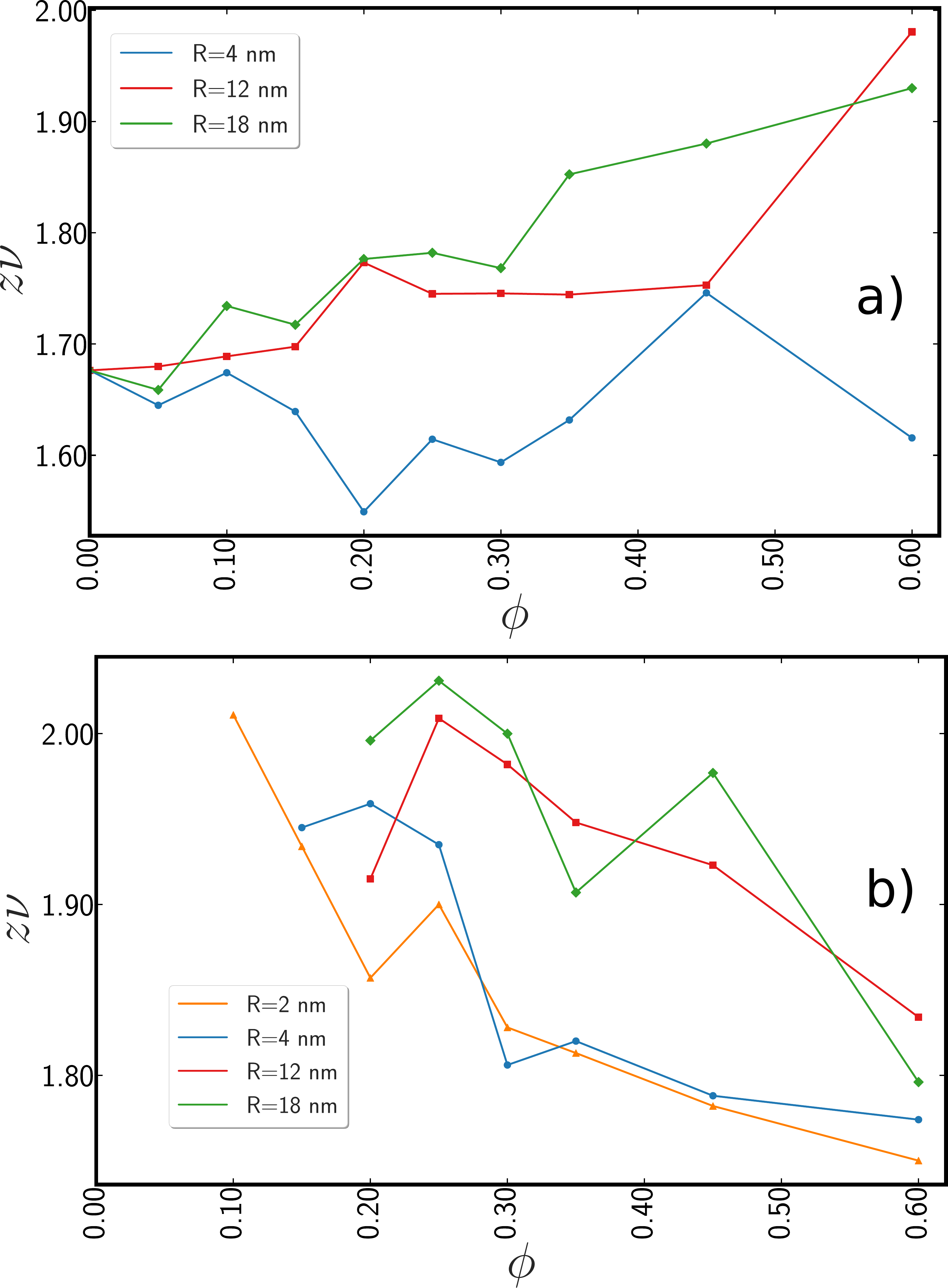}
		\caption{The exponent \(z \nu\) versus volume fraction for different colloid sizes for the a) LB, and b)Langevin simulations.}
		\label{z_nu}

\end{figure}

The fits to the correlation functions yield the relaxation time at different $p$, with the resulting shown in log–log plots in Figs.~\ref{tau_p_LG_96} for one example with the other cases shown in the supplementary material.   From Eq.(\ref{tau_scaling}), we see that the slope of these plots on a log-log plot yields the exponent \( z\nu \).  The results of these fits are shown in Figure~\ref{z_nu}.  While there is significant fluctuations in the results from statistical noise due to the limited runtimes possible, clear trends emerge from the data.

\begin{figure}[ht]
		\includegraphics[trim= 100 0 0 0,scale=0.15]{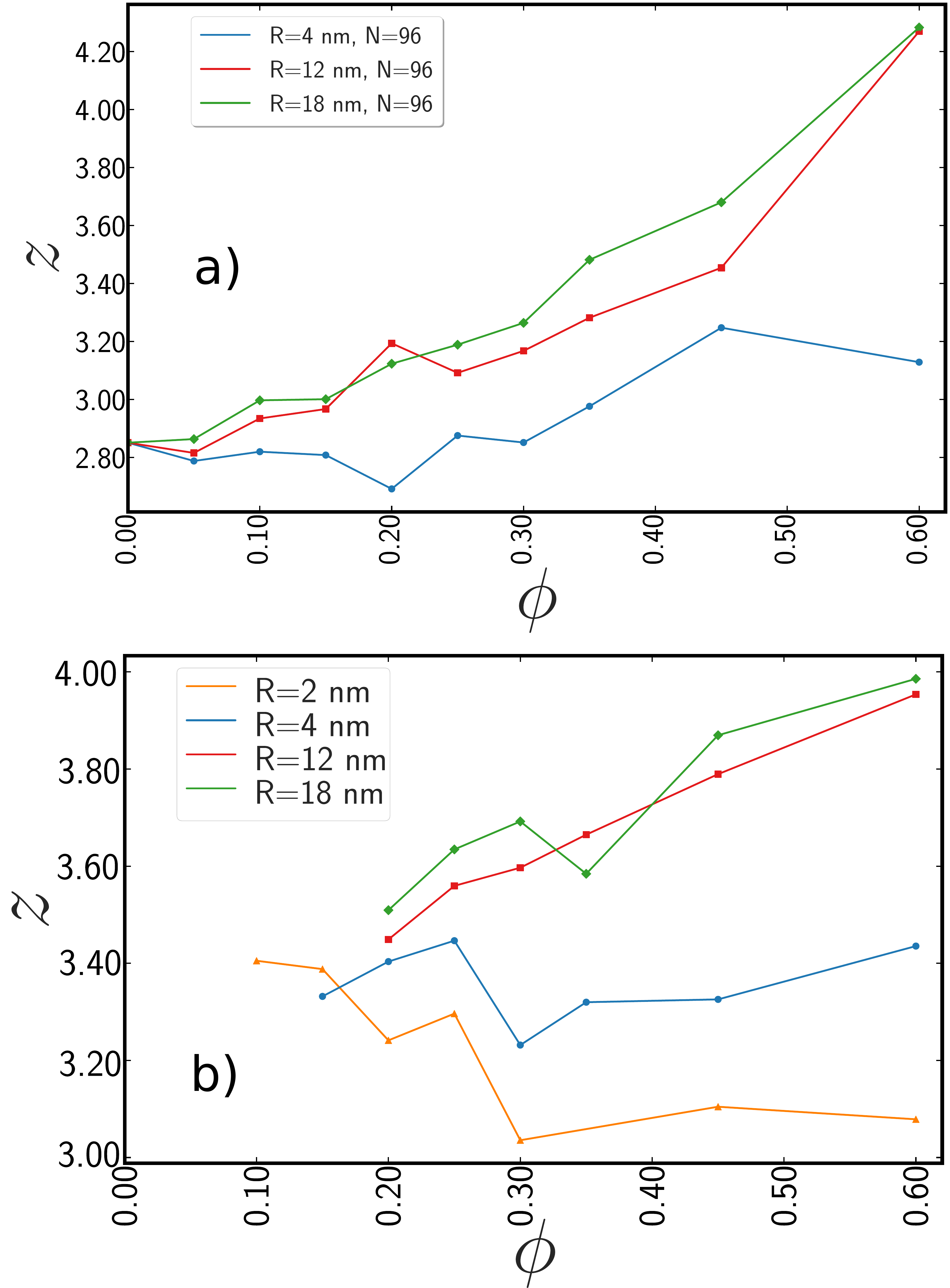}
		\caption{$z$ versus $\phi$ for different colloid sizes for a) LB, and b)Langevin simulations.}
		\label{z_vs_phi}

\end{figure}

To isolate the dynamic exponent $z$, we divide $z\nu$ by the Flory exponent $\nu$ found in Figure~\ref{nueff}. The resulting plot of $z$ versus $\phi$ reveals distinct regimes in Figure~\ref{z_vs_phi}. The LB polymer includes HI intrinsically, and as expected, it shows $z \approx 3$ at low to intermediate $\phi$. However, at high $\phi$ and for very large colloid sizes, the LB model shows a deviation, with $z$ increasing. This is likely an example of hydrodynamic screening, where the dense colloids with large sizes obstructs the fluid's long-range flow, causing the system to lose its Zimm-like character.  The smallest colloids show less significant deviation from $z \approx 3$ and only at higher volume fractions.  In this case, these colloids are effectively modifying the hydrodynamic solvent: as they are comparable in size to the monomers of the polymers interactions between monomers along the chain still have a hydrodynamic nature whereas the large colloids can completely block flow on scales comparable to $R_g$ damping the hydrodynamic interaction.  In contrast, for the LD polymer, which lacks explicit hydrodynamic interactions (HI), we observe that at moderate to high $\phi$ with colloids of comparable size to the monomers, $z$ approaches 3. This Zimm-like behavior suggests that the dense (small) colloid bath acts as an effective solvent, inducing HI on the length scale of the polymer (the Langevin thermostat should still damp out hydrodynamic modes at long wavelengths).  When the colloids are large, this mechanism breaks down and $z$ only increases as $phi$ increases, with values similar to the large colloid case for the LB simulations where the large colloids screen the hydrodynamic interactions.

\section{Discussion and Conclusions}

We have investigated how mobile colloidal crowders modify both the static structure and the dynamics of a polymer, focusing on the roles of crowder size, volume fraction $\phi$, and hydrodynamic interactions.

As a reference, a single polymer in pure solvent reproduces the expected limiting behavior. Static conformations follow good-solvent scaling, $R_g \sim N^{\nu}$ with $\nu \approx 0.59$, independent of the dynamical method. Dynamically, Langevin dynamics (LD) exhibits Rouse behavior with $D \sim 1/N$, while Lattice--Boltzmann (LB) simulations recover Zimm-like scaling, $D \sim N^{-\nu}$, reflecting solvent-mediated hydrodynamic coupling. These baseline results confirm that our simulations correctly capture the classical limits.

Upon introducing colloids, the polymer becomes progressively compact with increasing $\phi$. However, the mechanism of compaction depends strongly on colloid size. Small colloids primarily affect short-wavelength structure by enhancing local flexibility, while long-wavelength scaling remains close to good-solvent behavior. In contrast, large colloids reduce the effective Flory exponent at long length scales, indicating a degradation of effective solvent quality. Analysis of the static structure factor identifies a crossover wavevector corresponding to the mean monomer--colloid separation distance, consistent with a de Gennes blob picture: polymer segments behave as free chains inside voids, while at larger scales the chain becomes a random walk of confinement blobs. Larger colloids create larger cavities and stronger long-scale confinement, leading to more pronounced global compaction.

The diffusion behavior highlights clear differences between LD and LB. In the LD simulations, polymer diffusion decreases rapidly with increasing $\phi$ and shows a strong dependence on colloid size. In particular, small colloids produce a pronounced reduction in diffusion even at low volume fraction. A simple renormalization of the monomer friction, $D_0/D = \gamma_\phi/\gamma_0$, is insufficient to collapse the data. Introducing a phenomenological size-dependent factor $\ln(1 + R/r)$ provides a reasonable collapse across different colloid sizes, indicating that the effective drag depends not only on concentration but also on the geometric ratio between colloid radius $R$ and monomer size $r$. 

Scaling the LG diffusion data by $R_g$ further reduces scatter, suggesting that polymer-size effects also contribute. Although LD does not explicitly include hydrodynamic interactions, the non-overlapping nature of colloids means that motion of one colloid influences nearby colloids. At moderate densities, clusters of interacting colloids can generate correlated motion over length scales comparable to $R_g$, effectively mimicking hydrodynamic-like interactions at the polymer scale. This also explains why the smallest colloids at very low volume fraction ($\phi \lesssim 0.05$) deviate from the general trend: at low density they are too far apart to induce correlated motion, and the polymer experiences essentially independent obstacles.

In contrast, LB diffusion follows an exponential dependence similar to an effective-medium on the volume fraction over a wide range, $\eta_\phi/\eta_0 \sim \exp(b \phi)$, consistent with the interpretation of the suspension as a fluid of increased effective viscosity. The dependence on colloid size is much weaker than in LD, reflecting the dominance of solvent-mediated hydrodynamic coupling. At high volume fraction ($\phi \approx 0.6$), deviations from effective-medium scaling emerge as the system approaches random close packing. Colloids form long-lived cages, the confinement length becomes comparable to $R_g$, and polymer motion becomes cage-limited, proceeding through loop rearrangements and cage-to-cage transitions rather than simple reptation.

Rouse mode analysis provides a length--scale resolved view of the dynamics. The distinction between LD and LB is especially clear in this analysis. In LB, hydrodynamic interactions mediate long-range coupling between monomers, and all modes follow the expected scaling $\tau_p^{-1} \sim (p/N)^{z\nu}$, correlation functions are exponential from $t=0$, and scaling behavior is preserved.  In LD, at low $\phi$, scaling breaks down first at short wavelengths (large $p$), while long-wavelength modes remain closer to classical behavior. The breakdown is strongest when the volume fraction is low, the colloids are large, and short-wavelength modes are examined, reflecting sparse and heterogeneous polymer--colloid interactions.
In LD, this scaling recovery at moderate $\phi$ suggests that a dense colloid bath can induce effective Zimm-like behavior. However, at very high $\phi$ and for large colloids, particularly in LB, the dynamic exponent increases above 3, consistent with hydrodynamic screening. Dense, large colloids obstruct long-range fluid flow, and the system gradually loses its Zimm-like character.

Polymer transport in crowded suspensions results from a balance between steric blocking, geometric confinement effects, correlated motion of the obstacles, and hydrodynamic screening. The role of hydrodynamics in particular appears underappreciated, with a common assumption that it does not play an important role in confined systems.  However, using Langevin dynamics to model the colloidal suspensions can lead to erroneous conclusions about the impact of the crowder size on the polymer dynamics, which is significant in Langevin simulations but almost absent in simulations with full hydrodynamics. Our findings provide a clearer physical picture of how polymers behave in crowded soft-matter and biological environments.

\section{Supplementary Material}
Additional plots and data supporting the findings of this study that are referenced in the main text above are included in a supplementary material file.

\begin{acknowledgments}
We gratefully acknowledge financial support from the Natural Sciences and Engineering Research Council of Canada (NSERC, CD: RGPIN-2025-06298) and the Ontario Graduate Scholarship (OGS). This research was enabled by the use of computing resources provided by the Shared Hierarchical Academic Research Computing Network (SHARCNET) and the Digital Research Alliance of Canada.
\end{acknowledgments}

\section*{Data Availability Statement}
The data that support the findings of this study are available within the article and its supplementary material.

\section*{Author Declarations}
The authors have no conflicts to disclose.

\clearpage

	\bibliography{AIP_JCP_apssamp}

\end{document}